\documentclass[traditabstract]{aa} 

\usepackage[T1]{fontenc}
\usepackage{ae,aecompl}
\usepackage{mathrsfs}
\usepackage{hyperref}
\usepackage{txfonts}

\usepackage{graphicx}   
\usepackage{amsmath}    
\usepackage{amssymb}    

\usepackage{caption}
\usepackage{cuted}

\usepackage{lscape}

\usepackage{changepage}
\usepackage{mathtools}
\usepackage{booktabs}
\usepackage{adjustbox}
\usepackage{xcolor}
\usepackage{multirow}

\defcitealias{Lynch23}{LL23}

\title{Local spherical collapsing box in \textsc{Athena++}: Numerical implementation and benchmark tests}

   \author{Ziyan Xu
          \inst{1,2} \thanks{E-mail: ziyan.xu@ens-lyon.fr}
          \and
          Elliot M. Lynch\inst{1} \thanks{E-mail: elliot.lynch@ens-lyon.fr}
          \and
          Guillaume Laibe
          \inst{1}
          }

   \institute{$^{1}$ Univ Lyon, Univ Lyon 1, ENS de Lyon, CNRS, Centre de Recherche Astrophysique de Lyon UMR5574, F-69230, Saint-Genis-Laval, France\\
   $^{2}$ Center for Star and Planet Formation, GLOBE Institute, University of Copenhagen, Øster Voldgade 5–7, 1350 Copenhagen, Denmark
             }

\date{Accepted XXX. Received YYY; in original form ZZZ}

\begin{document}


\abstract{We implement a local model for a spherical collapsing or expanding gas cloud in the \textsc{Athena++} magnetohydrodynamic code.
This local model consists of a Cartesian periodic box with time-dependent geometry.
We present a series of benchmark test problems, including nonlinear solutions and linear perturbations of the local model, confirming the code's desired performance. During a spherical collapse, a horizontal shear flow is amplified, corresponding to angular momentum conservation of zonal flows in the global problem; wave speed and the amplitude of sound waves increase in the local frame, due to the reduction in the characteristic length scale of the box, which can lead to an anisotropic effective sound speed in the local box.
Our code conserves both mass and momentum-to-machine precision.
This numerical implementation of the local model has potential applications to the study of local physics and hydrodynamic instabilities during protostellar collapse, providing a powerful framework for better understanding the earliest stages of star and planet formation.}

   
\maketitle


\section{Introduction} \label{sec:intro}

Star formation has been studied over decades both theoretically and numerically  on large spatial scales, the key motivation being to determine how global redistribution of angular momentum results in the formation of stars, disks, or jets and set the initial conditions for planet formation (see \citealt{Larson69, Penston69} and e.g., \citealt{Pattle23,Pineda23,Tsukamoto23} for recent reviews). The field has been stimulated by instruments capable of resolving the largest scales of these young stellar systems in ever-finer detail in several wavelength domains, such as the Atacama Large Millimeter Array (ALMA), the Very Large Telescope (VLT) or the \textit{James Webb} Space Telescope (JWST).
The importance of these questions has perhaps overshadowed that of the physical processes taking place simultaneously on a small scale. One question, for example, is whether it is possible for dust to start concentrating as soon as the cloud collapses, thereby promoting highly efficient planet formation later on.

The local box formalism provides a simple treatment of the hydrodynamics of a spherically symmetric system undergoing contraction. The global effects of curvature and acceleration due to collapse are consistently incorporated into the local treatment of the problem by source terms that can be time-dependent, while allowing for the adoption of periodic boundaries as the system is spatially homogeneous. Numerical simulations using the local approximation approach play an important role in the study of local gas dynamics and instabilities in astrophysics. 
The best-known example is the rediscovery of the magnetorotational instability (MRI) with the local shear box approximation, which was first applied to magnetohydrodynamic (MHD) simulations by  \citet{Hawley95}. Local models have been used for (isotropic) cosmological collapses \citep{Robertson12}. Anisotropy in collapse and expansion was taken into account in the study of starless molecular core collapse  \citep{Toci18}, solar winds \citep{Velli92,Grappin93,Grappin96,Tenerani17,Shi20,Squire20} (among which \citet{Tenerani17} also consider nonuniform expansion), anisotropy-driven plasma kinetic instabilities \citep{Sironi15,Bott21}, and the cosmic ray pressure anisotropy instability \citep{Sun23}.

Recently, \citet{Lynch23} (hereafter \citetalias{Lynch23}) developed a new local model for the spherical collapse or expansion problem, allowing for a nonuniform collapse or expansion. This model considers a radially varying flow, as in \citet{Tenerani17}, but also allows for the time-dependence of the background flow. It also uses a different treatment of pressure gradient than the expanding box models, in an approach that is more similar to the distorted shearing box models of \citet{ogilvie13a,Ogilvie14}. This avoids spurious instabilities from the inclusion of formally subdominant terms \citep{Latter17}, allowing for the application of periodic or shear-periodic boundary conditions with a Cartesian-like geometry. In this study, we incorporate \citetalias{Lynch23}'s model into \textsc{Athena++}, a high-order Godunov MHD code \citep{stone20}. Other local models have already been implemented in \textsc{Athena++}, including \citet{Squire20} for the (uniformly) expanding box, and MHD-PIC \citep{Sun23}, also allowing for an anisotropic collapse or expansion. However, most of these previous numerical implementations focus on the study of MHD properties and magnetized waves, in the context of stellar winds or cosmic rays, and the hydrodynamic properties of the expanding or collapsing box have not been comprehensively tested so far. Here, we aim to incorporate the local model of \citetalias{Lynch23}, with benchmark tests to thoroughly investigate and validate the hydrodynamic properties therein.

This article is organized as follows. We first describe the formulation of the local model for the collapsing box in Sect. \ref{sec:equations}. In Sect. \ref{sec:numerical}, we describe the numerical implementation, especially the modification of the Riemann solver in order to incorporate the local model. We then conduct a series of test cases including nonlinear solutions and linear perturbations in Sect. \ref{sec:tests}, as well as numerical studies. We further discuss possible applications and perspectives, and give future perspectives in Sect. \ref{sec:discussion}. We summarize and conclude in Sect. \ref{sec:conclusion}.

\begin{figure*}
\begin{center}

\includegraphics[width=\textwidth]{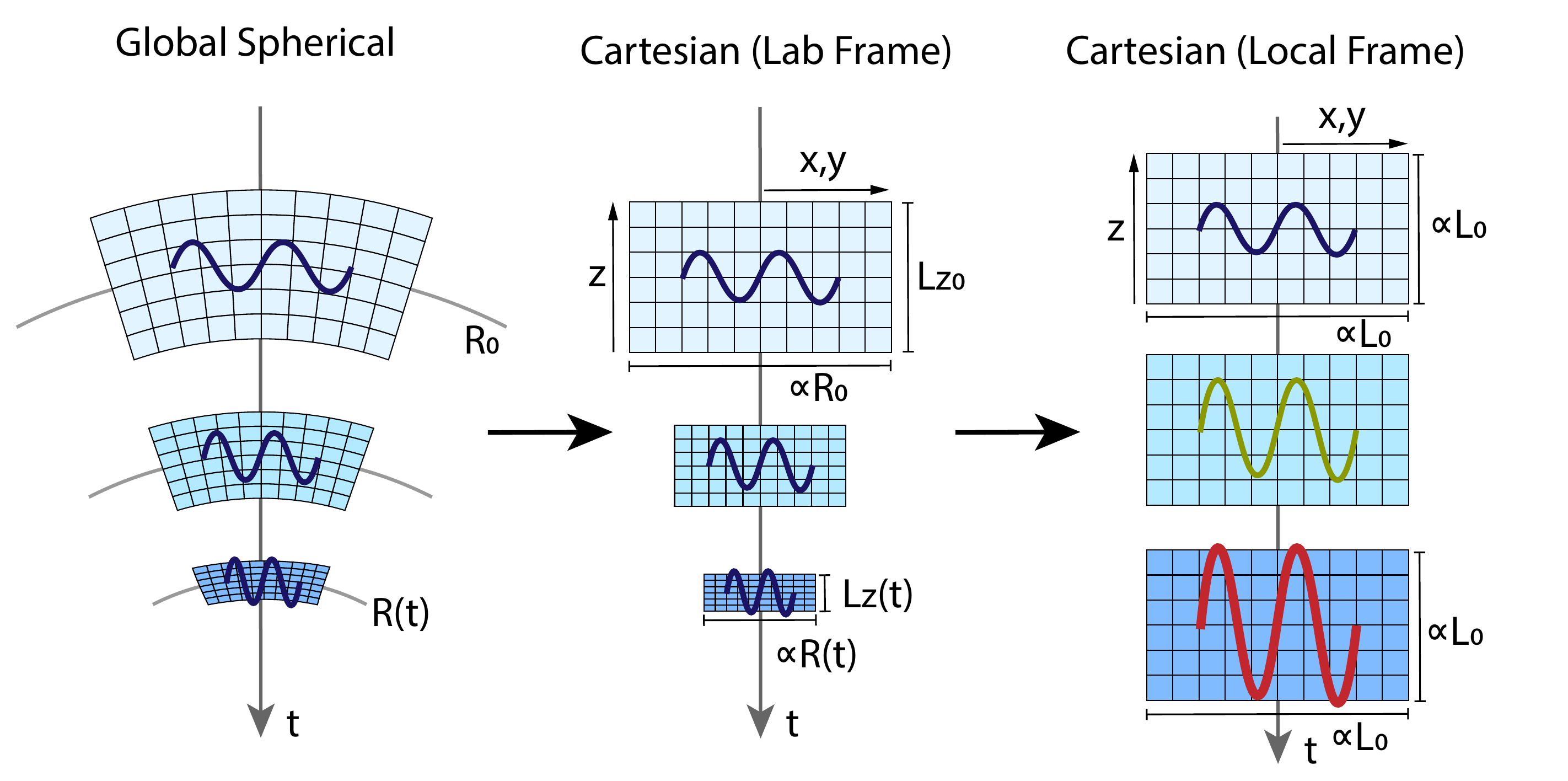}
\end{center}
\caption{Cartoon illustration of the local collapsing box model. On the left, the global geometry depicts a domain moving radially due to background flow. The volume of the domain varies as it approaches the center of the spherical cloud, leading to changes in density. A compressing domain leads to a density increase, as is shown by the progressively darker blue color in the sequence. 
The horizontal domain size is proportional to $\mathcal{R}\left(t \right)$, establishing the horizontal length scale in the local model (middle panel). The vertical length scale, $L_z \left( t\right)$, is determined by the distance between the domain's inner and outer radial boundaries, which is set by the background flow. 
Inside each box, a waveform signal illustrates how certain physical quantities evolve in the local model. The width of the waveform represents quantities that remains fixed in the local frame (e.g., wavelength). The amplitude of the waveform represents specific physical quantities that are expected to be constant in the global geometry throughout the collapse (e.g., angular momentum, isothermal physical sound speed). In the local reference frame (right), the corresponding quantities such as the shear flow velocity or the effective sound speed are amplified as the collapse proceeds and the length scale shrinks in the domain, while quantities such as the wavelength remain the same. This phenomenon is visually demonstrated by the vertically stretched waveforms in the sequence. For a more accurate and detailed understanding of this process, one should refer to the equations in \S \ref{sec:equations} .\label{fig:illustration}}
\end{figure*}

\section{Equations of the local model} \label{sec:equations}

We started by considering an axisymmetric spherical gas cloud collapsing or expanding with a purely radial velocity, $U_{bg}(r,t)$, that varies with both radius, $r$, and time, $t$. We then considered a local neighborhood of a reference point located at a radius, $\mathcal{R}(t)$, in the spherical gas cloud, described by a local Cartesian coordinate system comoving with the background flow. The local coordinates are given by
\begin{equation}\label{eq:coord}
x=\phi, y=\theta-\frac{\pi}{2}, z=\frac{r-\mathcal{R}(t)}{L_z(t)},
\end{equation} 
where $\phi$ and $\theta$ are the longitude and latitude on the reference sphere of radius, $\mathcal{R}$, and $r-\mathcal{R}(t)$ is the local radial distance relative to the global, $\mathcal{R}(t)$, of the reference frame. We note that $x$, $y$, $z$ have dimensions of angles in this formalism, and that they denote two azimuthal and radial directions, respectively. $L_z$ is a characteristic vertical length scale that is set by the background flow profile, which can be approximated by
\begin{equation}
U_{bg}(r,t) = U_0(t)+U_{R0}(t)\left(r - \mathcal{R}(t)\right).
\end{equation}
The Lagrangian time derivative in the local model is
\begin{equation}
D = \partial_t + v_i \partial_i,
\end{equation}
while the divergence of the velocity of the radial background flow is 
\begin{equation}
\Delta = \frac{2U_0}{\mathcal{R}}+U_{R0} .
\end{equation}

The flow within the local box model represents local, nonlinear perturbations to the background flow. Following the derivation in Sect. 2 of \citetalias{Lynch23}, in the local frame, the continuity, momentum, and energy equations are given in the stretched Cartesian coordinate system by

\begin{align}
D \rho  &= - \rho \left[ \Delta + \partial_x v_{x} + \partial_y v_{y} + \partial_{{z}} v_{{z}} \right], \label{eq:mass}\\ 
D v_{x} + \frac{2 U_0}{\mathcal{R}} v_{x} &= - \frac{1}{\rho \mathcal{R}^{2}} \partial_{x} p , \label{eq:momx}\\
D v_{y} + \frac{2 U_0}{\mathcal{R}} v_{y} &= - \frac{1}{\rho \mathcal{R}^{2}} \partial_{y} p \label{eq:momy}, \\
D v_{{z}} + 2 v_{{z}} U_{R 0} &= - \frac{1}{\rho L_z^2} \partial_{{z}} p \label{eq:momz} , 
\end{align}
\begin{align}
\begin{split}
&\partial_t\left(\rho \mathcal{E}_{\text {rell }}\right)+\partial_i\left[ v_i\left(\rho \mathcal{E}_{\text {rell }}+p\right)\right]= \\
& - \rho\left[U_R L_z^2 v_z^2+\frac{U_0}{\mathcal{R}} \mathcal{R}^2\left(v_x^2+v_y^2\right)\right]- \left(p +\rho \mathcal{E}_{\text {rell }}\right) \Delta,
 \end{split}
 \label{eq:energy}
\end{align}
where $\rho$ is the density of the gas, $v_i$ is the relative velocity field in the local frame,\footnote{The velocities here are the contravariant components of the velocity. In this article, we use subscript indices to denote the spatial dimensions (e.g., $v_i$) for convenience, as opposed to superscript indices (e.g., $v^i$) as is conventionally used. We also apply Einstein summation throughout this paper, where repeated indices (subscript or superscript) are implicitly summed over.} $p$ is the pressure of the gas, and $\mathcal{E}_{rell}$  is the energy density of the relative motion in this system, (re-)defined as
\begin{equation}
\mathcal{E}_{\text {rell }}=\frac{1}{2} L_z^2 v_z^2 +\frac{1}{2} \mathcal{R}^2(t)\left(v_x^2+v_y^2\right)+\varepsilon
,\end{equation}
which is the sum of the kinetic energy density of relative motion in the local frame, $\mathcal{E}_{k,\text {rell }} \equiv  \frac{1}{2} L_z^2 v_z^2 +\frac{1}{2} \mathcal{R}^2(t)\left(v_x^2+v_y^2\right)$, and the specific internal energy, $\varepsilon$, determined by the equation of state (EoS). For an isothermal EoS, $p = \rho c_s^2$, where $c_s$ is the sound speed, and the energy equation (\ref{eq:energy}) is dropped from the system.  For an adiabatic EoS, $\varepsilon=\frac{p}{\rho(\gamma-1)}$ , where $\gamma$ is the ratio of specific heats, assumed to be constant. 

The right-hand side of Eqs.~\ref{eq:momx} -- \ref{eq:momz} can be interpreted as an anisotropic sound speed in the local coordinate. We define two rescaled sound speeds in the horizontal and vertical directions separately: $c_{\mathrm{s},\mathrm{H}} \equiv c_{\mathrm{s}} /\mathcal{R} $ and $c_{\mathrm{s},\mathrm{V}} \equiv c_{\mathrm{s}} / L_z$.

Eqs.~\ref{eq:mass} -- \ref{eq:energy} can be written in a compact form,
\begin{equation}\label{eq:compact}
\frac{\partial \mathbf{q}}{\partial t}+\frac{\partial \mathbf{f}}{\partial x}+\frac{\partial \mathbf{g}}{\partial y}+\frac{\partial \mathbf{h}}{\partial z}=\mathbf{S}\left(x,y,z,t \right),
\end{equation}
where the vector, $\mathbf{q}$, of conservative variables is defined according to
\begin{equation}
\mathbf{q} = \left(\begin{array}{c}
q_1  \\
q_2 \\
q_3\\
q_4 \\
q_5
\end{array}\right) \equiv \left(\begin{array}{c}
\rho  \\
\rho v_x \\
\rho v_y\\
\rho v_z \\
\rho \mathcal{E}_{\text {rell }}
\end{array}\right),
\end{equation}
$\mathbf{f}, \mathbf{g}$, and $\mathbf{h}$ are the fluxes of the associated conservative variables in the $x$, $y$, and $z$ directions, respectively:
\begin{align}
&\mathbf{f} \equiv \left(\begin{array}{c}
\rho v_x \\
\rho v_x^2+p/\mathcal{R}^2 \\
\rho v_x v_y\\
\rho v_x v_z \\
\left(\rho \mathcal{E}_{\text {rell }}+p\right) v_x
\end{array}\right),
\mathbf{g}\equiv\left(\begin{array}{c}
\rho v_y \\
\rho v_y v_x \\
\rho v_y^2+p/\mathcal{R}^2\\
\rho v_y v_z \\
\left(\rho \mathcal{E}_{\text {rell }}+p\right) v_y
\end{array}\right), \\
&\mathbf{h}\equiv\left(\begin{array}{c}
\rho v_z \\
\rho v_z v_x \\
\rho v_z v_y\\
\rho v_z^2+p/L_z^2\\
\left(\rho \mathcal{E}_{\text {rell }}+p\right) v_z
\end{array}\right).
\end{align}
The vector of source terms, denoted as $\mathbf{S}$, is
\begin{equation} \label{eq:sourceterm}
\mathbf{S}\equiv\left(\begin{array}{c}
-\rho \Delta \\
-2\rho v_x\frac{U_0}{\mathcal{R}} -\rho \Delta v_x  \\
-2\rho v_y\frac{U_0}{\mathcal{R}} -\rho \Delta v_y \\
-2\rho v_z U_{R0} -\rho \Delta v_z \\
- \rho\left[U_R L_z^2 v_z^2+\frac{U_0}{\mathcal{R}} \mathcal{R}^2\left(v_x^2+v_y^2\right)\right]- \left(p +\rho \mathcal{E}_{\text {rell }}\right) \Delta
\end{array}\right).
\end{equation}
Again, for an isothermal EoS, the fifth component of the vectors $\mathbf{q}$, $\mathbf{f}$, $\mathbf{g}$, $\mathbf{h}$, and $\mathbf{S}$ are dropped, as the energy equation is not solved for this case.

A more intuitive illustration of the local model geometry is presented in Fig. \ref{fig:illustration}. As the collapse proceeds, the local domain is compressed in volume, leading to an increase in gas density. This is reflected in the source term in the continuity equation (\ref{eq:mass}). The angular momentum conservation law leads to amplification of horizontal flow velocity in the local frame during the collapse, which is represented by the source terms in the momentum equations on the horizontal directions (Eqs. \ref{eq:momx} and \ref{eq:momy}).  These features along with other characteristic behaviors of the local model are further illustrated and tested in Sect. \ref{sec:tests}.

\section{Numerical implementation}
\label{sec:numerical}

We next translated the local model of \S~\ref{sec:equations} into a Godunov scheme, and implemented it in the numerical code \textsc{Athena++}. Finite volume discretization of the compact form of the hydrodynamic equations Eq.(\ref{eq:compact}) gives the time integration step as a set of ordinary differential equations,

\begin{equation}
\begin{aligned}
\mathbf{q}_{i,j,k}^{n+1}=\mathbf{q}_{i,j,k}^n&-\frac{\delta t}{\delta x}\left(\mathbf{f}_{i+1 / 2,j,k}^{n+1 / 2}-\mathbf{f}_{i-1 / 2,j,k}^{n+1 / 2}\right)\\
&-\frac{\delta t}{\delta y}\left(\mathbf{g}_{i,j+1/2,k}^{n+1 / 2}-\mathbf{g}_{i,j-1/2,k}^{n+1 / 2}\right)\\
&-\frac{\delta t}{\delta z}\left(\mathbf{h}_{i,j,k+1 / 2}^{n+1 / 2}-\mathbf{h}_{i,j,k-1 / 2}^{n+1 / 2}\right)\\
&+\delta t \mathbf{S}_{i,j,k}^n\\
\equiv \mathbf{\tilde{q}}_{i,j,k}^{n}
+\delta t \mathbf{S}_{i,j,k}^n,
\end{aligned}
\end{equation}

where subscript indices $(i,j,k)$ denote numerical cells centered at positions $(x_i,y_j,z_k)$, superscripts, $n$, denotes the index for discrete time, $\delta x$, $\delta y$, and $\delta z$ are discrete grid sizes, and $\delta t$ is the time step. The vectors $\mathbf{q}_{i,j,k}^{n}$ and $\mathbf{S}_{i,j,k}^n$  correspond to volume-averaged $\mathbf{q}$ and $\mathbf{S}$  over the numerical cells. The vectors $\mathbf{f}_{i+1 / 2,j,k}^{n+1 / 2}$, $\mathbf{g}_{i+1 / 2,j,k}^{n+1 / 2}$, and $\mathbf{h}_{i+1 / 2,j,k}^{n+1 / 2}$ correspond to time-averaged $\mathbf{f}$, $\mathbf{g}$, and $\mathbf{h}$, respectively. The half-integer subscripts on the flux vectors refer to the edges of computational cells, while the half-integer superscripts denote the time average between $t^n$ and $t^{n+1}$.  $\mathbf{\tilde{q}}_{i,j,k}^{n}$ denotes the vector of conservative variables after flux integration.

Except for the pressure gradient term of the momentum equation, which was treated during the calculation of numerical fluxes in \textsc{Athena++}, all source terms in the continuity, momentum, and energy equations can be added explicitly with an operator-splitting method, following the standard \textsc{Athena++} approach. However, examining the source term vector in eq. (\ref{eq:sourceterm}), all of the components in the source term, $\mathbf{S}$, can be integrated analytically. Thus, instead of following the standard explicit approach, we incorporated the source terms by updating the conservative variables with the analytical solutions at $t^{n+1}$; namely,
\begin{align}
\mathbf{q}_{i,j,k}^{n+1}= \mathbf{\psi}\left( \mathbf{\tilde{q}}_{i,j,k}^{n} \right),
\end{align}
where 
\begin{align}
\mathbf{\psi}\left( \mathbf{q} \right)=\left(\begin{array}{c}
\psi_1  \\
\psi_2\\
\psi_3\\
\psi_4\\
\psi_5 
\end{array}\right) =\left(\begin{array}{c}
\frac{J^n}{J^{n+1}} ~ q_1 \\
\left(\frac{\mathcal{R}^n}{\mathcal{R}^{n+1}}\right)^2 \frac{J^n}{J^{n+1}} ~ q_2\\
\left(\frac{\mathcal{R}^n}{\mathcal{R}^{n+1}}\right)^2 \frac{J^n}{J^{n+1}} ~ q_3\\
\left(\frac{L_z^n}{L_z^{n+1}}\right)^2 \frac{J^n}{J^{n+1}} ~ q_4\\
\psi_1 \mathcal{E}_{k,\text {rell }}^{n+1} + q_1 \varepsilon \left( \frac{J^n}{J^{n+1}}\right)^{\gamma}
\end{array}\right),
\end{align}
with 
$J^n \equiv \mathcal{R}^n \mathcal{R}^n L_z^n$ the Jacobian determinant in the local coordinate system, and 
\begin{align}
\mathcal{E}_{k,\text {rell }}^{n+1}  =\frac{1}{2 \psi_1^2}\left(L_z^{n+1}\right)^2  {\psi_4^2} + \frac{1}{2 \psi_1^2}\left(\mathcal{R}^{n+1}\right)^2 \left({\psi_2^2}+{\psi_3^2}\right) 
\end{align}
the kinetic energy density of relative motion at $t^{n+1}$. This approach of analytical integration of the source term significantly enhances the numerical accuracy, and ensures that the conservation laws are fully obeyed (see Sect. \ref{subsec:conservation}). The method we used to integrate the source term is first-order in time. While higher-order methods have been explored, they are not as efficient within the current framework. We further discuss the impact of the treatment of the source term on numerical accuracy in Sect. \ref{subsec:convergence}.

The flux integration (or the time-averaged numerical fluxes) mentioned above were computed depending on different time-integrating methods \citep[see \S 3.2.3 of][for details]{stone20}. To explain the core of the algorithm, we present hereafter the flux calculation in the $x$ direction only as an example, dropping the subscripts $j$ and $k$ for simplicity. For a time integrator with a stage number of $N_{stage}$ in each time step, for each stage $s=0,...N_{stage-1}$, $\mathbf{q}_i^n$ was updated with fluxes $\mathbf{f}_{i-1 / 2}^s$ and $\mathbf{f}_{i-1 / 2}^s$, which were computed with a Riemann solver. 
In \textsc{Athena++}, various solvers were implemented, including Roe's linearized solver \citep{Roe81}, as well as the Harten-Lax-van Leer (HLL) approximated Riemann solvers \citep{Harten83}. These approximate Riemann solvers provide an estimate of the fluxes. As was discussed above, pressure gradient terms in the momentum equation were integrated during the calculation of fluxes in \textsc{Athena++}. Since the corresponding terms in the local model (the right-hand side of Eqs.~\ref{eq:momx} -- \ref{eq:momz}) can be interpreted as an anisotropic sound speed,  we modified the Riemann solver, which incorporates the anisotropic effective sound speed by rescaling the wave speed during flux calculation. We applied this modification in Roe's solver, the Harten-Lax-van Leer-Einfeldt (HLLE) solver, and the Harten-Lax-van Leer-Contact (HLLC) solver, the latter being recommended for adiabatic hydrodynamics. The basic introduction of these Riemann solvers can be found in \citet{Toro99}, so will not be introduced in detail.

 For the flux, $\mathbf{f}_{i-1 / 2}^s$, at the cell interface, $x_{i-1 / 2}$,  the Roe's solver and HLLE solver provide the estimated flux, $\mathcal{F}_{i-1 / 2}^{\mathrm{Roe}}$ and $\mathcal{F}_{i-1 / 2}^{\mathrm{HLLE}}$, separately. The Roe's flux is
\begin{equation}
\mathcal{F}_{i-1 / 2}^{\mathrm{Roe}}=\frac{1}{2}\left(\mathbf{f}_{L, i-1 / 2}+\mathbf{f}_{R, i-1 / 2}+\sum_\alpha a^\alpha\left|\lambda^\alpha\right| \mathscr{R}^\alpha\right).
\end{equation}
The HLLE flux at the interface, $x_{i-1 / 2}$, is
\begin{equation}
\mathcal{F}_{i-1 / 2}^{\mathrm{HLLE}}=\frac{b^{+} \mathbf{f}_{L, i-1 / 2}-b^{-} \mathbf{f}_{R, i-1 / 2}}{b^{+}-b^{-}}+\frac{b^{+} b^{-}}{b^{+}-b^{-}}\left(\mathbf{q}_i-\mathbf{q}_{i-1}\right).
\end{equation}
In both cases, the subscripts $L$ and $R$ denote the left and right states of each variable at the interface, and $\mathbf{f}_{L, i-1 / 2}=\mathbf{f}\left(\mathbf{q}_{L, i-1 / 2}\right)$ and $\mathbf{f}_{R, i-1 / 2}=\mathbf{f}\left(\mathbf{q}_{R, i-1 / 2}\right)$ are the fluxes evaluated using the left and right states of the conserved variables. 

For Roe's solver, $\lambda^\alpha$ denotes the $\alpha$-th eigenvalue of the Roe's matrix in the conserved variables (see e.g. $\S 4.3 .2$ and Appendix B of \citep{Stone08}), and 
\begin{equation}
a^\alpha=\mathscr{L}^\alpha \cdot \left(\mathbf{q}_{L, i-1 / 2}-\mathbf{q}_{R, i-1 / 2}\right),
\end{equation}
where the $\mathscr{L}^\alpha$ and $\mathscr{R}^\alpha$ are the $\alpha$-th left and right eigenvectors of Roe's matrix, which corresponds to $\lambda^\alpha$. $\mathscr{L}^\alpha$ and $\mathscr{R}^\alpha$ are also rows and columns of the left and right eigenmatrices, $\mathscr{L}$ and  $\mathscr{R}$. For isothermal hydrodynamics, the eigenvalues of Roe's matrix are
\begin{align}
    \lambda=\left(v_x-C, v_x, v_x, v_x+C\right) \text {, }
\end{align}

where $C$ is the (effective) sound speed. The left and right eigenmatrices are
\begin{align}
    \mathscr{L}=\left[\begin{array}{cccc}
\left(1+v_x / C\right) / 2 & -1 /(2 C) & 0 & 0 \\
-v_y & 0 & 1 & 0 \\
-v_z & 0 & 0 & 1 \\
\left(1-v_x / C\right) / 2 & 1 /(2 C) & 0 & 0
\end{array}\right]
\end{align}

and
\begin{align}
    \mathscr{R}=\left[\begin{array}{cccc}
1 & 0 & 0 & 1  \\
v_x-C & 0 & 0 & v_x+C  \\
v_y & 1 & 0 & v_y  \\
v_z & 0 & 1 & v_z  \\
\end{array}\right].
\end{align}

\begin{figure}[htb!]
\begin{center}
\includegraphics[width=0.51\textwidth]{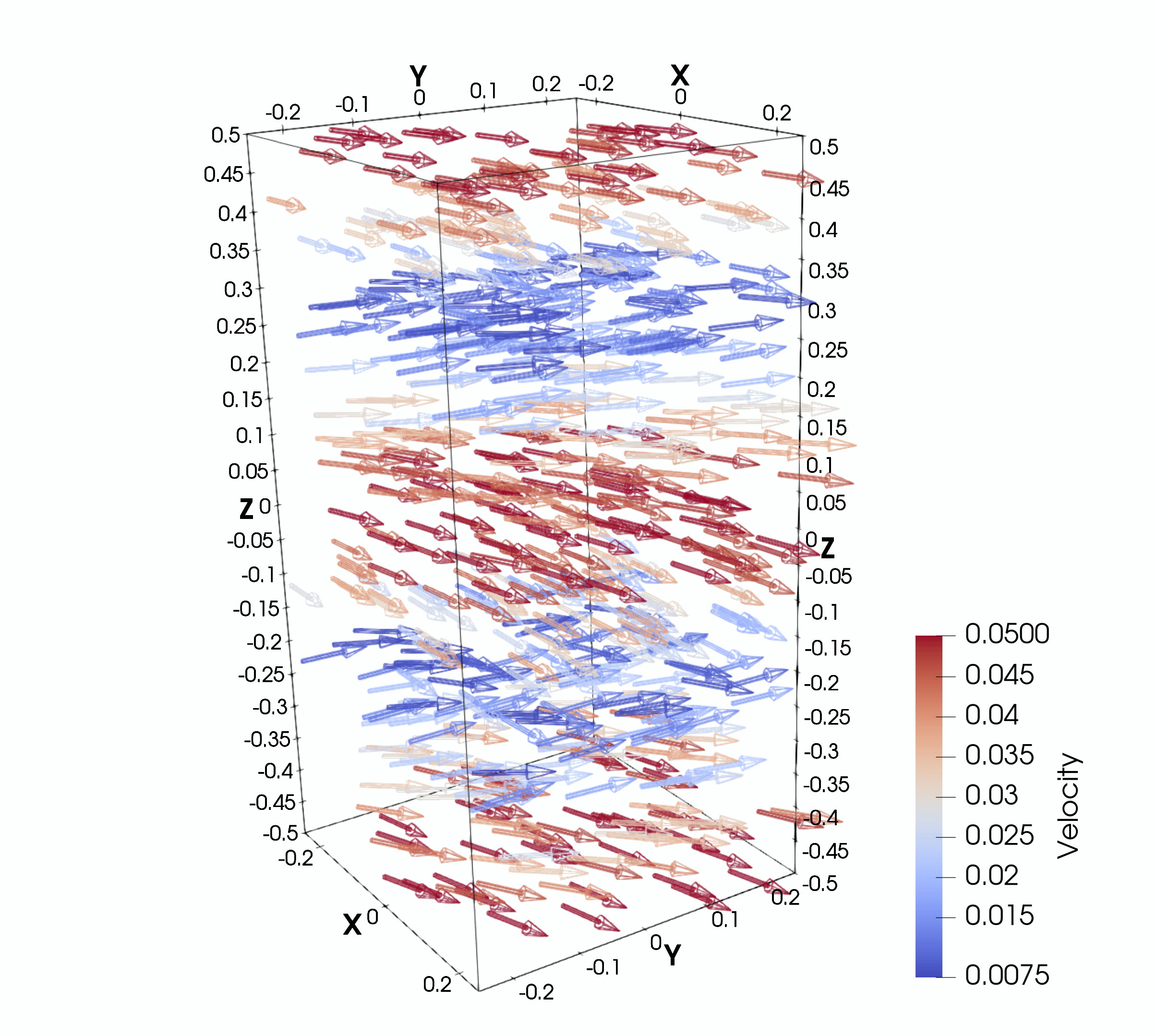}
\includegraphics[width=0.43\textwidth]{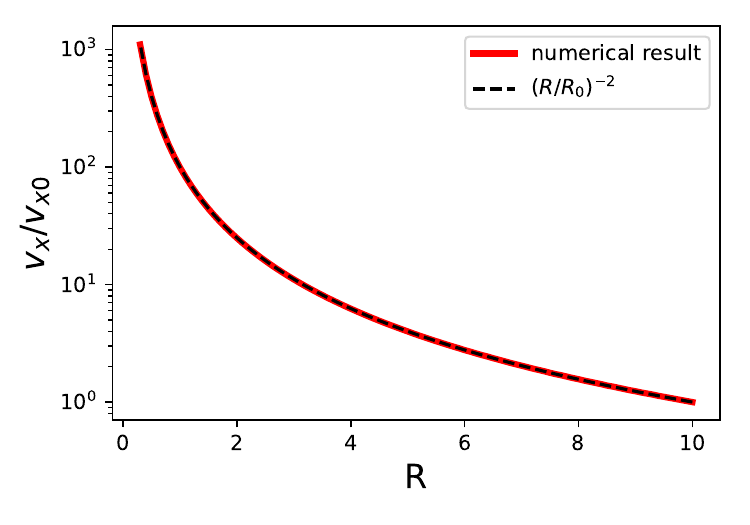}
\includegraphics[width=0.43\textwidth]{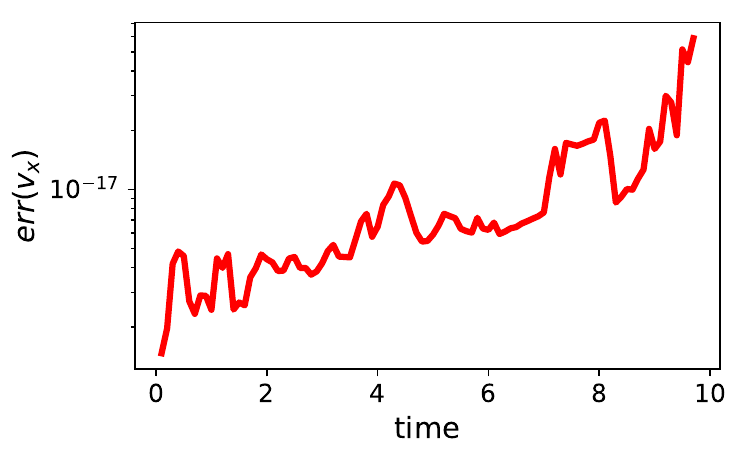}
\end{center}
\caption{Test case of horizontal shear flows. 
Upper panel: 3D visualization of the initial spatial distribution of the shear flow in our simulation. Arrows mark the direction of shear flow, and the total flow velocity magnitude is mapped to colors (see the color bar). Middle panels: Solid red line shows  $v_x/v_{x0}$ vs. R, i.e., normalized shear velocity as a function of global radius of the local box, showing growth of shear velocity as collapse proceeds, closely matches the analytical expectation (dashed black lines). The result of $v_y$ is identical to that of $v_x$ and is not shown here. Lower panel: Time evolution of relative error $\mathrm{err}(v_x)$. \label{fig:shear}}
\end{figure}

Here, we rescaled the effective sound speed with a scaling factor, $l$, so that $C = c_{\mathrm{s}}/l$, with $c_{\mathrm{s}}$ the isothermal or adiabatic sound speed. In the $x$ and $y$ directions, we chose $l=\mathcal{R}$ so that $C=c_{\mathrm{s},\mathrm{H}}$; in the $z$ direction, $l=L_z$ so we have $C=c_{\mathrm{s},\mathrm{V}}$. We note that when $l=1$, we returned to the original eigenvalues and eigenvectors given in \cite{Stone08}. 
For adiabatic hydrodynamics, the eigenvalues and the eigenmatrice are a bit more complicated than the isothermal case as the (modified) energy equation is also solved. The corresponding quantities are shown in App.~\ref{sec:adiabatic} .

Similarly, for the HLLE solver, 
\begin{align}
    \begin{gathered}
b^{+}=\max \left[\max \left(\lambda^M, v_{x, \mathrm{R}}+C_{\mathrm{R}}\right), 0\right] , \\
b^{-}=\min \left[\min \left(\lambda^0, v_{x, \mathrm{L}}-C_{\mathrm{L}}\right), 0\right] .
\end{gathered}
\end{align}

Here, $\lambda^M$ and $\lambda^0$ denote the maximum and minimum eigenvalues of Roe's matrix given above, $v_{x, \mathrm{L}}$ and $v_{x, \mathrm{R}}$ are the velocity component normal to the interface in the left and right states, respectively, and $C_{\mathrm{L}}=c_{\mathrm{s},\mathrm{L}}/l$ , $C_{\mathrm{R}}=c_{\mathrm{s},\mathrm{R}}/l$ are the effective sound speed computed from the left and right states. For the HLLC solver, the middle contact wave is also included. 
To compute the fastest wave speed including the middle wave, the rescaled sound speeds, $C_{\mathrm{L}}$ and $C_{\mathrm{R}}$, were also used to estimate the pressure in the intermediate states. In the middle wave speed estimation processes, the pressure terms, $p^*$, that originate from the momentum equations were replaced with $p^*/l^2$. 

The spatial reconstruction makes use of the left and right eigenmatrices of the Roe's matrix in the primitive variables, and was modified accordingly. For isothermal hydrodynamics, the eigenmatrices are identical to the original ones presented in Appendix B of \citep{Stone08}, with the isothermal sound speed replaced by the effective sound speed, $C$, in our case, as was illustrated previously. For adiabatic hydrodynamics, the eigenmatrices are again in a different form, and are shown in  App.~\ref{sec:adiabatic}. We implemented this modification for both piecewise linear (PLM) and piecewise parabolic (PPM) spatial reconstructions.

Rescaling the sound speed implies the modification of the Courant–Friedrichs–Lewy (CFL) condition. We further considered the effect of the time-varying sound speed on the stability condition, by applying the leapfrog finite difference equation (FDE) to a harmonic oscillator with a time-varying frequency. Our new CFL condition is thus given by
\begin{align}
 \delta t = C_{\circ} \cdot \min \left[\frac{\delta x ~\mathrm{v}(\beta_x )}{|v_{x}|+c_{\mathrm{s},\mathrm{H}}} ,\frac{\delta y ~\mathrm{v}(\beta_y) }{|v_{y}|+c_{\mathrm{s},\mathrm{H}}},\frac{\delta z ~\mathrm{v}(\beta_z) }{|v_{z}|+c_{\mathrm{s},\mathrm{V}}}\right],
 \label{eq:CFL}
\end{align}
where $C_{\circ}$ is the CFL number,\footnote{Time discretization in \textsc{Athena++} was performed following the method of lines, for which the stability condition depends on both the number of dimensions and the integration method. In a 1D problem, or in multi-dimensional problems with a strong-stability preserving integrator \citep[e.g. second- and third- order Runge-Kutta,][]{Gottlieb09}, the stability condition is generally $C_{\circ}\leq 1$. However, for a 2D or 3D van Leer integrator, the condition is $C_{\circ}\leq 1/2$.} and the original physical sound speed has been replaced by the effective sound speeds, $c_{\mathrm{s},\mathrm{H}}$ and $c_{\mathrm{s},\mathrm{V}}$. The effect of the time-varying sound speed is reflected in $\mathrm{v}(\beta_x )$, $\mathrm{v}(\beta_y )$, and $\mathrm{v}(\beta_z )$, with $\beta_x = \beta_y = -U_0/c_{\mathrm{s}}$, and $\beta_z = -U_{R0} L_z /c_{\mathrm{s}}$. Both $\beta$ and $\mathrm{v}(\beta)$ are dimensionless. For a collapsing system ($\beta > 0$), which is the focus of this work, we have
\begin{equation}
    \mathrm{v}(\beta) = \frac{\sqrt{1+8\beta}-1}{4\beta}.
\end{equation}
For a static system, $\beta=0$, we have $\mathrm{v}(\beta \to 0)=1$, which goes back to the original CFL condition, but with rescaled sound speeds. The case of the expanding system ($\beta < 0$), as well as the derivation of the new CFL condition, are described in App. \ref{sec:cfl}.

This local model is physically meaningful with periodic boundary conditions. Further extension of the shear-periodic boundary condition (local model with weak rotation) will be considered in future work.

\section{Numerical tests}
\label{sec:tests}

In this section, we present benchmark tests of the local box for nonlinear solutions (Sect. \ref{subsec:flows}) and linear waves (Sect. \ref{subsec:wave}).  Fiducially, we used Roe's solver with the van Leer integrator (VL2) and PPM spatial reconstruction for our showcase tests, with an isothermal EoS. Other parameters of simulation setups, including simulation box sizes and resolutions, are described in each subsection (Sects. \ref{subsec:flows}, \ref{subsec:wave}, and \ref{subsec:conservation}) for each individual tests.  Additionally, we present convergence studies in Sect. \ref{subsec:convergence}, examining both Roe and HLL solvers, considering various time integrators including VL2, second- and third-order Runge-Kutta (RK2, RK3) integrators, with both PPM and PLM reconstructions, and both isothermal and adiabatic EoSs (with $\gamma=1.4$) tested. The CFL number was set to $C_{\circ}=0.4$ for shear flow tests (Sect. \ref{subsec:flows}) and $C_{\circ}=0.2$ for wave tests (Sect. \ref{subsec:wave}).
We set the code unit with an initial background (unperturbed) gas density of $\rho_0=1$ and a physical sound speed of $c_{s0}=1$. The length unit of the code, $L_0=1$, was set by the unit of the local coordinate system defined in Eq. (\ref{eq:coord}) (see also the right panel of Fig. \ref{fig:illustration}). We also defined the initial values of the length scales in the local model, $R_0 \equiv \mathcal{R}(0)$, $L_{z0} \equiv L_z(0)$, for convenience.
For each numerical test, we compared our numerical results with analytical solutions to validate the accuracy and robustness of our implementation. We briefly summarize the analytical solutions corresponding to the tests in each subsections separately. Details of the analytical solutions are presented in \citetalias{Lynch23}.

\subsection{Nonlinear solutions} \label{subsec:flows}

\begin{figure}[htb!]
\begin{center}
\includegraphics[width=0.45\textwidth]{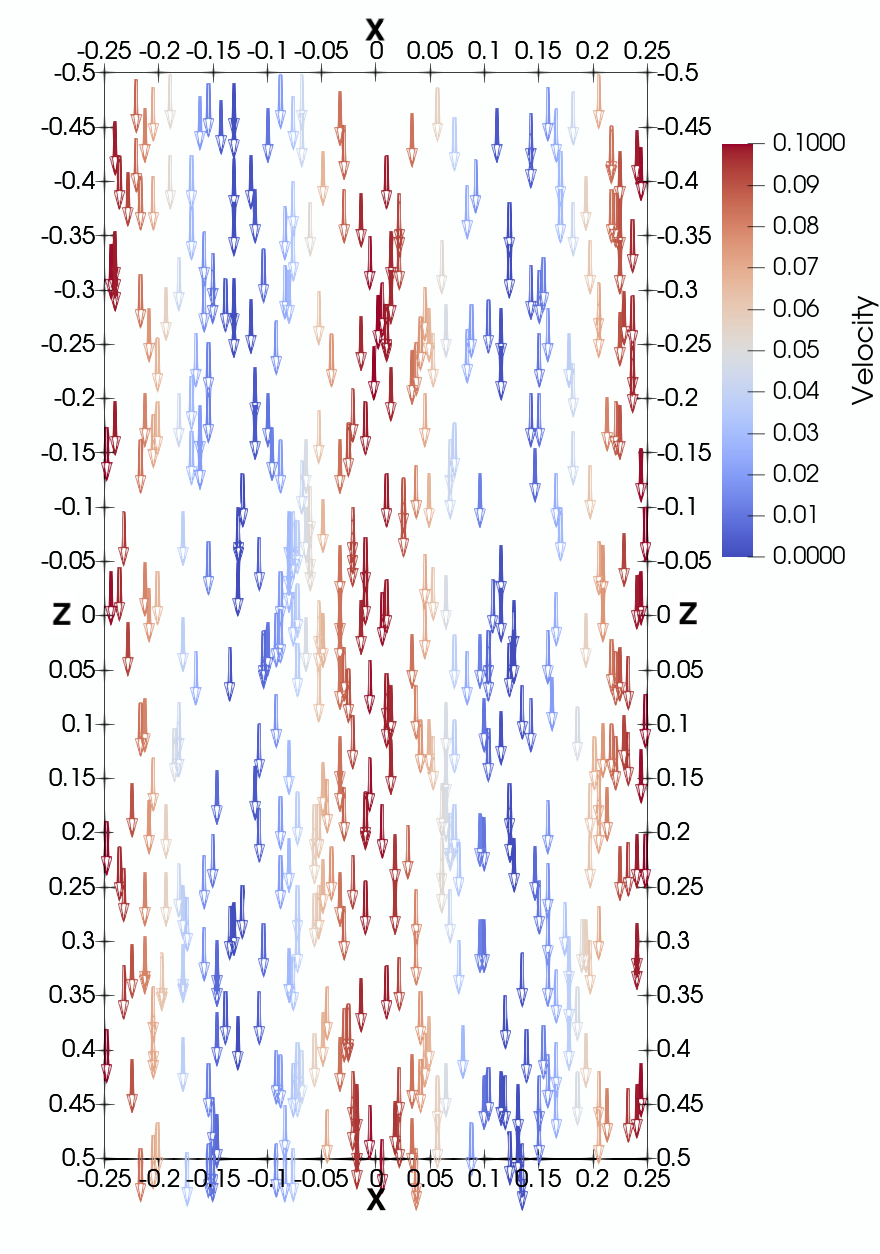}
\includegraphics[width=0.4\textwidth]{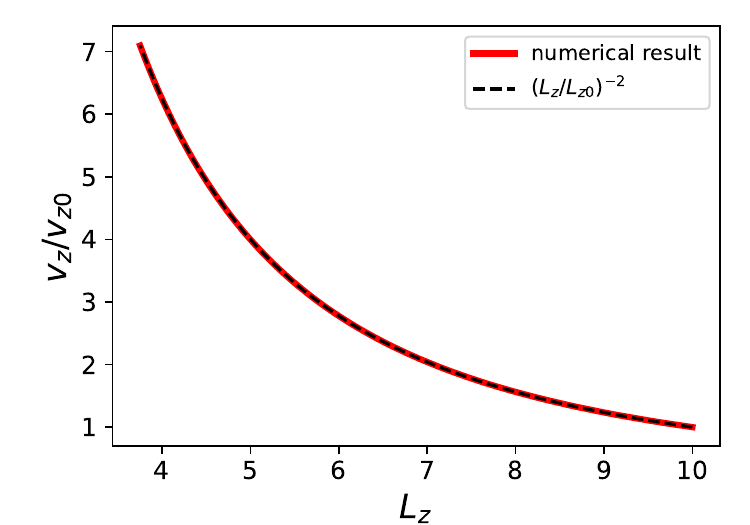}
\includegraphics[width=0.4\textwidth]{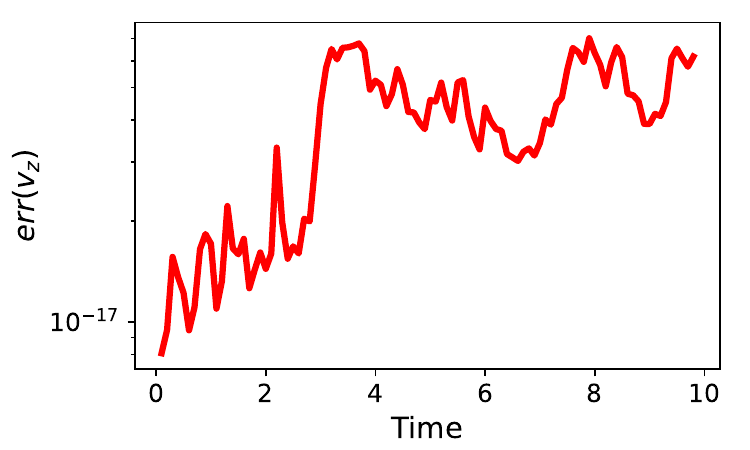}
\end{center}
\caption{Same as Fig. \ref{fig:shear}, but for the test case of elevator flows. \\
The vertical flow velocity grows as the simulation box collapses, closely matching the analytical expectation.  \label{fig:elevator}}
\end{figure}
\subsubsection{Horizontal shear flows} \label{subsec:3dshear}

Consider a local model with a spatially homogeneous density, $\rho = \rho(t)$, and a purely horizontal velocity field, 
\begin{align}
 &v_{x} = A(t) v_{x0} (1+\cos(k_z z))  , \\
 &v_{y} = A(t) v_{y0} (1+\sin(k_z z))  , \\
&v_{z} = 0.
\end{align}
This is a horizontal shear flow that varies only in the vertical direction. Without loss of generality, we can set $A(0)=1$, and the solution to this local model is given with the evolution of the flow amplitude, $A$,
\begin{equation}
A = \left( \mathcal{R} \left( t \right)/R_0  \right)^{-2},
\end{equation}
yielding the following velocity field at time, $t$,
\begin{align}
 &v_{x} = v_{x0}(1+\cos(k_z z))   \left(\mathcal{R} \left( t \right)/R_0\right)^{-2}, \\
 &v_{y} = v_{y0}(1+\sin(k_z z))  \left(\mathcal{R} \left( t \right)/R_0\right)^{-2}.
\end{align}
In global geometry, this growth in the horizontal flow velocity is a byproduct of the conservation of angular momentum during the spherical collapse.

\begin{figure*}[htb!]
\begin{center}
\includegraphics[width=\textwidth]{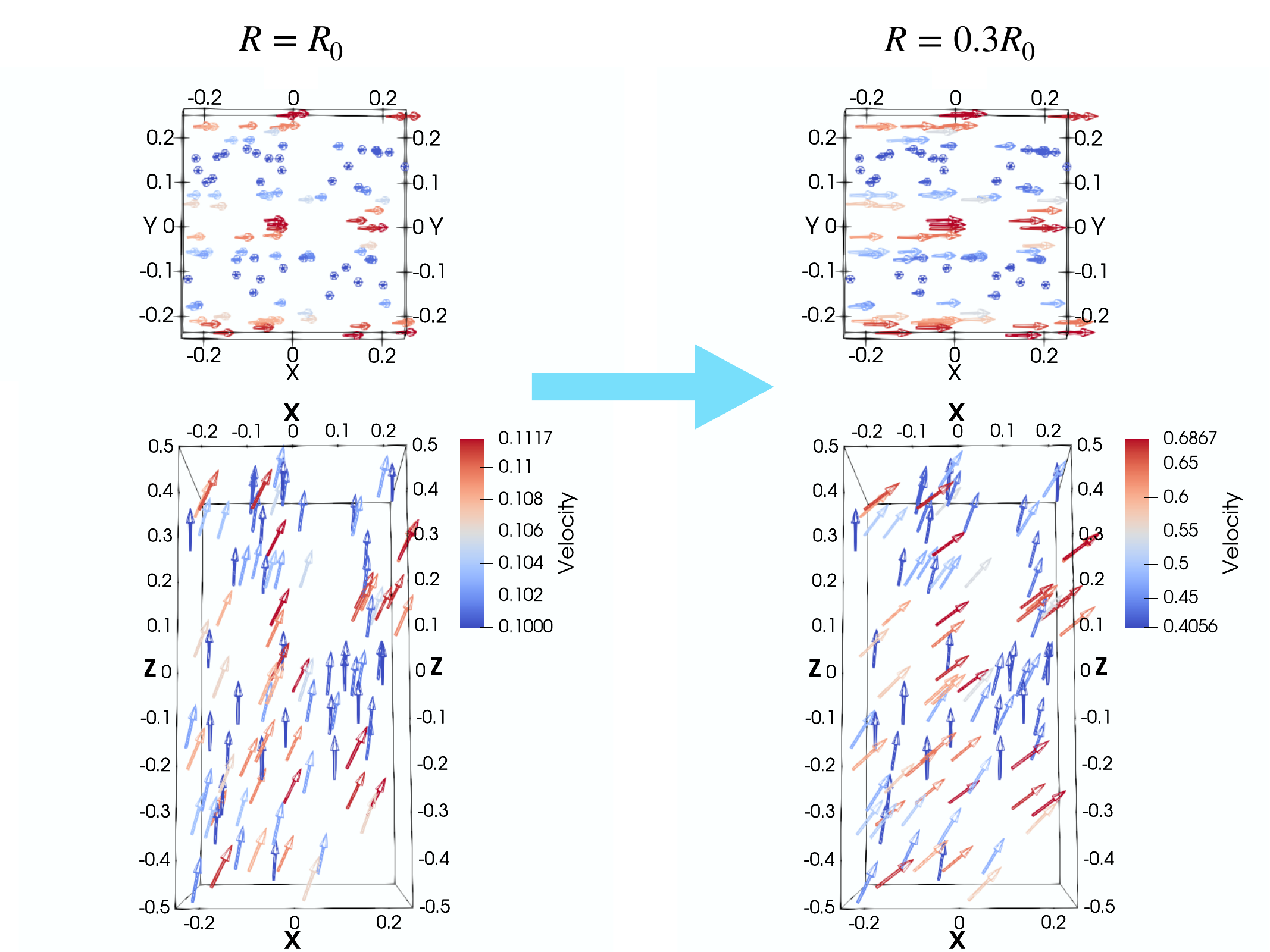}
\end{center}
\caption{
3D visualization of the spatial distribution of the diagonal flow test. Arrows mark the direction of the flow, and the total flow velocity magnitude is mapped to colors (see the color bar). 
Left panels are the initial setup ($\mathcal{R}=R_0$); right panels are at $\mathcal{R}=0.3R_0$. Flow direction varies as the collapse proceeds. \label{fig:diagflow}}
\end{figure*}

We tested this solution by setting a local model of uniform collapse, with  $R_0 = 10.0 L_0$, $U_0 = -1.0 c_{\mathrm{s}0}$, $ L_{z0}=10.0L_0$, and $U_{R0}=0.0$. The initial flow structure was set with $v_{x0}=0.025c_{\mathrm{s}0}$, $v_{y0}=0.0125c_{\mathrm{s}0}$, and $k_z=4\pi$. The simulation box size was set with $L_{x,\mathrm{sim}}=L_{y,\mathrm{sim}}=0.5L_0, L_{z,\mathrm{sim}}=1.0L_0$, with a resolution of 128 cells per unit length. We note that our choice of simulation box size quickly becomes too large to be physically meaningful; however, this is irrelevant to numerically testing the local model. The initial distribution of this velocity field is shown in the upper panel of Fig. \ref{fig:shear}, presenting a 3D visualization of the simulation box. Arrows mark the direction of the velocity vectors, while colors mark the velocity magnitude.

The middle panel of Fig. \ref{fig:shear} present the simulation result with the peak shear velocity as a function of $\mathcal{R}$, where the peak shear velocity is measured by the maximum value of the horizontally averaged flow velocity. As was expected, the shear flow is amplified as the collapse proceeds, in excellent agreement with the analytical solution. We further defined the relative error with L2 norm of each variable, $\mathcal{X}$,
\begin{align} \label{eq:err}
\mathrm{err}\left(\mathcal{X}\right) = \frac{1}{N}\sqrt{\sum_i \left(\frac{\mathcal{X}_i - \mathcal{X}_{i,\mathrm{exact}}}{\mathcal{X}_{max,\mathrm{exact}}}\right)^2},
\end{align}
where $N$ is the total cell number, $i$ denotes each cell, and $\mathcal{X}_i$ and $\mathcal{X}_{i,\mathrm{exact}}$ are the numerical result and analytical solution for $\mathcal{X}$, respectively. We used the maximum absolute value of the analytical solution over the domain, $\mathcal{X}_{max,\mathrm{exact}} = \max(\left| \mathcal{X}_{i,\mathrm{exact}} \right|)$, to normalize the relative error, in order to avoid meaninglessly large relative errors when the analytical solution is zero or a small number. The time evolution of $\mathrm{err}(v_{x})$ is displayed in the lower panel of Fig. \ref{fig:shear}, showing that our result matches the analytical solution to machine precision.

\subsubsection{Elevator flows}\label{subsec:elevator_flow}

We considered a second class of nonlinear flows in the local model, with a spatially homogeneous density, $\rho=\rho(t)$, and a purely vertical velocity field, 
\begin{align}
&v_{x} = v_{y} = 0,\\
&v_{z} = B (t) v_{z0} (1+\cos(k_x x))  .
 \label{elevator flow}
\end{align}
This vertical “elevator flow” has a solution with a temporally evolving flow amplitude, 
\begin{equation}
 B = (L_z/L_{z0})^{-2} ,
\end{equation}
if without loss of generality we set $B(0)=1$. This leads to a flow velocity of
\begin{equation}
 v_{z} =  v_{z0} (1+\cos(k_x x)) (L_z/R_0)^{-2} .
\end{equation}
These elevator flows can only occur with vertically periodic boundary condition, and could be an artifact of periodic boundaries.

To test the solution of these elevator flows, we set the local model with $R_0=10.0L_0$, $U_0=0.0$, $L_{z0}=10.0L_0$, and $U_{R0}=-0.1c_{\mathrm{s}0}L_0$. The flow structure was initialized with $v_{z0}=0.05c_{\mathrm{s}0}$ and $k_x=8\pi$. We chose a 2D simulation box with $L_{x,\mathrm{sim}}=0.5L_0$ and $L_{z,\mathrm{sim}}=1.0L_0$, and a resolution of 256 cells per unit length. This initial flow structure is displayed in the upper panel of Fig. \ref{fig:elevator}.

The middle panel of Fig. \ref{fig:elevator} presents the simulation result of the peak flow velocity as a function of $L_z$. The peak flow velocity was obtained by the maximum value of the vertically averaged flow velocity. The elevator flow grows with time, closely matching the analytical expectation. We further measured the relative error, $\mathrm{err}(v_z)$, defined in Eq. (\ref{eq:err}). The lower panel of Fig. \ref{fig:elevator} of the time evolution of $\mathrm{err}(v_z)$ shows that our result matches the analytical solution to machine precision.

\begin{figure}[htb!]
\begin{center}
\includegraphics[width=0.45\textwidth]{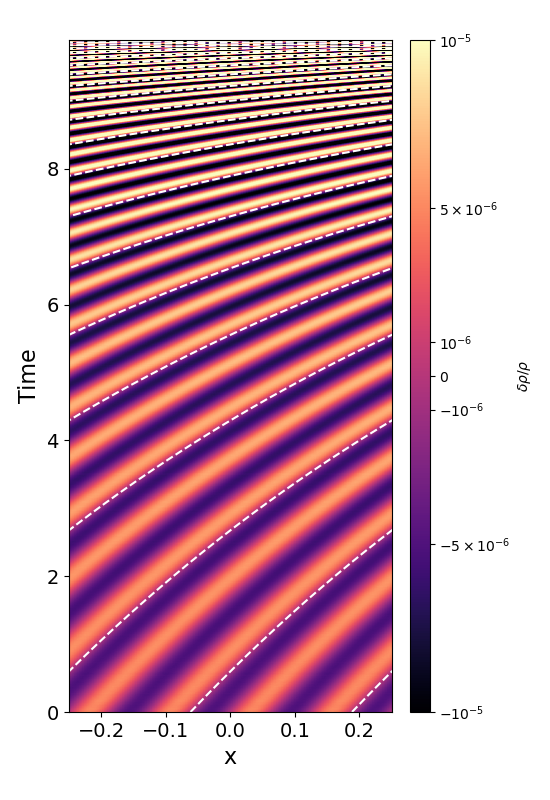}
\includegraphics[width=0.45\textwidth]{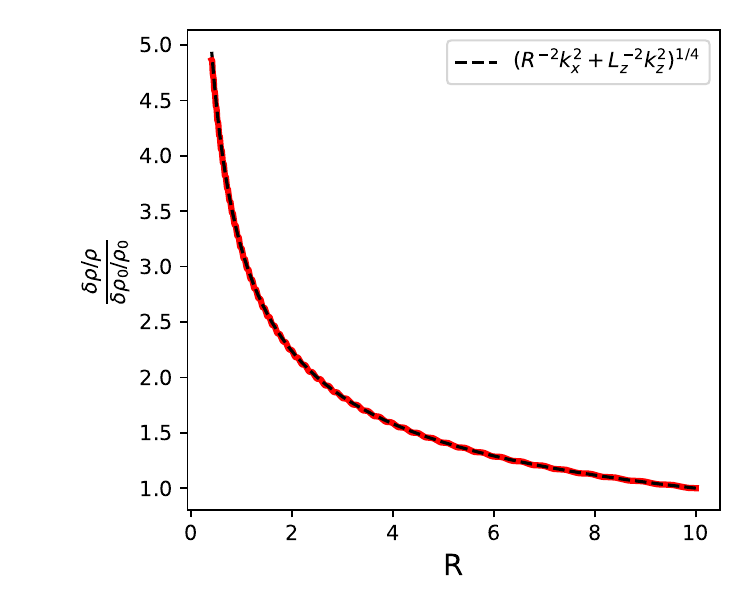}
\end{center}
\caption{\label{fig:soundwave_h}Test case of purely horizontal sound wave with uniform collapse. Upper panel: Space-time plot for vertically averaged relative density variation. Dashed lines show the theoretical expectation for phase evolution, in good agreement with the numerical results. Lower panel: Relative amplitude of perturbation, $\delta \rho/\rho$, as a function of R from the simulation (solid red line). The wave operates under the WKB regime, and is closely consistent with the theoretical expectation from the WKB approximation solution (dashed black line). The wiggles in the simulation results are not present in the WKB solution, but are consistent with the exact solution presented in Eq. (\ref{eq:exact_rho}).  } 
\end{figure}

\subsubsection{Diagonal flows} \label{diagonal flows}
We next considered a class of local models with a spatially homogeneous density of $\rho=\rho(t)$ and nonzero horizontal and vertical velocities. The solution for these diagonal flows combines $A=\left(\mathcal{R}/R_0\right)^{-2}$ of the horizontal shear flows and $B=\left(L_z/R_0\right)^{-2}$ of the elevator flows such that
\begin{align}
 v_x &= \left(\mathcal{R}/R_0\right)^{-2} v_{x0} (1+\cos(k_y y)) , \\
 v_z &= \left(L_{z}/L_{z0}\right)^{-2} v_{z0} (1+\cos(k_y y)) ,
\end{align}
where, without loss of generality, we have aligned the diagonal flow with the $x$ axis. In the global picture, these diagonal flows can be interpreted as a spiral with the pitch angle evolving over time as the collapse proceeds, since the horizontal and vertical components of the flow velocity change independently.

We tested this solution by setting up a local model with $R_0=10.0L_0$, $U_0=-1.0c_{\mathrm{s}0}$, $L_{z0}=10.0L_0$, and $U_{R0}=-0.1c_{\mathrm{s}0}/L_0$, with an initial flow structure of $v_{x0}=0.025c_{\mathrm{s}0}$, $v_{z0}=0.05c_{\mathrm{s}0}$, and $k_y=8\pi$. The simulation box was set with a box size of $L_{x,\mathrm{sim}}=L_{y,\mathrm{sim}}=0.5L_0$, $L_{z,\mathrm{sim}}=1.0L_0$, and a resolution of 128 cells per unit length. Figure \ref{fig:diagflow} presents the flow structure at the beginning of the simulation ($\mathcal{R}=R_0$), as well as later on at $\mathcal{R}=0.3R_0$. Variation in the flow direction is observed as the collapse proceeds. As $U_0>U_{R0}L_z$,  the horizontal flow component grows faster than the vertical component, causing the flow orientation to evolve toward the horizontal direction. The time evolution of flow amplitude closely matches the analytical solution in both the horizontal and vertical directions. The results are largely similar to those shown in Sect. \ref{subsec:3dshear} and Sect. \ref{subsec:elevator_flow}, and are not presented here. 

\subsection{Linear perturbation (waves) } \label{subsec:wave}
We examined the properties of linear perturbations in the local model (for details of the linear wave solutions refer to \S 4 and Appendix B of  \citetalias{Lynch23}; these are briefly summarized here). The density and velocity perturbations, $\delta \rho$ and $\delta \mathbf{v} $, of a diagonally propagating sound wave can be written as
\begin{align}
\delta v_x & =-\frac{2 c_\mathrm{s}}{\mathcal{R}^2}  k_x \Re\left[X \exp \left(i \mathbf{k} \cdot \mathbf{x}\right)\right], \\
\delta v_{{z}} & =-\frac{2 c_\mathrm{s}}{L_z^2}  k_z \Re\left[X \exp \left(i \mathbf{k} \cdot \mathbf{x}\right)\right], \\
\delta \rho & = \frac{2 \rho}{c_\mathrm{s}} \Im\left[\Pi \exp \left(i \mathbf{k} \cdot \mathbf{x}\right)\right], \label{eq:wave_rho}
\end{align}
where the (complex) wave amplitudes $(X,\Pi)$ fulfill
\begin{align} \label{eq:hamilton1}
\dot{X} &= \Pi, \\
\dot{\Pi} &= -\omega^2 X.\label{eq:hamilton2}
\end{align}
Here, we have introduced the sound wave frequency 
\begin{equation} \label{eq:omega}
\omega = c_\mathrm{s} |\mathbf{k}|=c_\mathrm{s}( \mathcal{R}^{-2}k_x^2+L_z^2k_{{z}}^2)^{1/2}.
\end{equation}
Without loss of generality, we aligned the wave propagating direction with the $x-z$ plane so that $\delta v_y = 0$.

The behavior of the wave is characterised by two regimes, depending on the value of the ratio between the wave period and the evolution timescale of the background flow, $t_{\mathrm{bg}}\sim \min\left(\mathcal{R}/\dot{\mathcal{R}}, L_z/\dot{L}_z\right)$. For a slowly varying background flow with $\omega t_{\mathrm{bg}} \gg 1$, the wave is described well by a Wentzel–Kramers–Brillouin (WKB) approximation. The WKB approximation for the density and velocity perturbations, $\delta \rho$ and $\delta \mathbf{v} $, of a diagonally propagating sound wave is
\begin{align}
\delta v_x & \approx-\frac{2 c_\mathrm{s}}{\mathcal{R}^2} \omega^{-1 / 2} k_x \Re\left[X_{\pm} \exp \left(i \mathbf{k} \cdot \mathbf{x}\pm i \int \omega d t\right)\right], \\
\delta v_{{z}} & \approx-\frac{2 c_s}{L_z^2} \omega^{-1 / 2} k_z \Re\left[X_{\pm} \exp \left(i \mathbf{k} \cdot \mathbf{x}\pm i \int \omega d t\right)\right], \\
\delta \rho & \approx \frac{2 \rho}{c_\mathrm{s}} \omega^{1 / 2} \Re\left[X_{\pm} \exp \left(i \mathbf{k} \cdot \mathbf{x}\pm i \int \omega d t\right)\right],\label{eq:wkb_rho}
\end{align}
where $X_{\pm}$ is some (complex) constant.

For a more general test, the exact solution for the sound waves can be obtained for a given collapse profile. One defines a length scale of reference  $L=L(t)$, and adopts
\begin{equation}
L=L_0\frac{\mathcal{R}}{\mathcal{R}_0}\left( \frac{k_x^2+b_0^{-2}k_z^{2}}{k_x^2+b^{-2}k_z^{2}}\right),
\end{equation}
where $b=L_z/\mathcal{R}$ is the aspect ratio of the local reference frame, and the subscription ``0'' denotes the initial values. This length scale of reference can be equal to the global radius of the cloud, $L=\mathcal{R}$, which can be achieved by adopting $L_0=\mathcal{R}_0$, and which can have either $b=b_0$ (constant aspect ratio) or $k_z=0$ (purely horizontal waves).
If we consider a collapse profile in the form of ${L} = L_0 \left( 1-t/t_\mathrm{c} \right)^\beta$, where $t_\mathrm{c}$ denotes the collapse timescale, this profile includes two typical types of collapse profiles; namely, uniform collapse (${L} = L_0 + U_0t$) where $\beta=1$, and a free-fall profile where $\beta=2/3$.  For uniform collapse, the exact solution for density and velocity is
\begin{align}
\delta v_x & =\frac{2 c_s}{\mathcal{R}^2}  k_x \sqrt{L} \Re\left[A_{\pm}L^{\pm \Lambda} \exp \left(i \mathbf{k} \cdot \mathbf{x}\right)\right], \\
\delta v_{{z}} & =\frac{2 c_s}{L_z^2} k_z \sqrt{L} \Re\left[A_{\pm} L^{\pm \Lambda} \exp \left(i \mathbf{k} \cdot \mathbf{x}\right)\right], \\
\delta \rho & = \frac{2 \rho L_0}{c_s^2t_c} \sqrt{L} \left\{\pm \Im\left[\left(\Lambda \pm 1/2\right)A_{\pm} L^{\pm \Lambda} \exp \left(i \mathbf{k} \cdot \mathbf{x}\right)\right]\right\}.\label{eq:exact_rho}
\end{align}
Here, we have introduced $\Lambda = \frac{1}{2}\sqrt{1-4\left(c_s|\mathbf{k}_0|t_c\right)^2}$, and $\left|\mathbf{k}_0\right|$ is the initial amplitude of $\mathbf{k}$.

For a general power law profile with $\beta \neq 1$, the exact solution can be written as
\begin{equation} \label{eq:powexact1}
\begin{aligned} 
\delta v_x  =\frac{2 k_x}{\mathcal{R}^2} s^{1 /[2(1-\beta)]} &\{\Re[A \exp (i \mathbf{k} \cdot \mathbf{x})]J_\nu(s) \\
&+\Re[B \exp (i \mathbf{k} \cdot \mathbf{x})] Y_\nu(s)\}, \\ 
\end{aligned}
\end{equation}
\begin{equation} \label{eq:powexact2}
\begin{aligned} 
\delta v_{{z}}  =\frac{2 k_{{z}}}{L_z^2} s^{1 /[2(1-\beta)]} &\{\Re[A \exp (i \mathbf{k} \cdot \mathbf{x})] J_\nu(s) \\
&+\Re[B \exp (i \mathbf{k} \cdot \mathbf{x})] Y_\nu(s)\}, \\ 
\end{aligned}
\end{equation}
\begin{equation} \label{eq:powexact3}
\begin{aligned} 
\delta \rho =-\frac{2 \rho}{c_s^2} \frac{1-\beta}{t_0} s^{\frac{1-2 \beta}{2(1-\beta)}}& \operatorname{sgn}(\dot{s}) \{\Im[A \exp (i \mathbf{k} \cdot \mathbf{x})] J_{\nu-1}(s) \\
&+\Im[B \exp (i \mathbf{k} \cdot \mathbf{x})] Y_{\nu-1}(s)\},
\end{aligned}
\end{equation}
where we have defined
\begin{equation} \label{eq:sandt0}
s = \left(\frac{t_c-t}{t_0}\right)^{1-\beta}, t_0 = \left[ \frac{|1-\beta|}{c_s|\mathbf{k}_0|t_c^{\beta}}\right]^{1/(1-\beta)}, 
\end{equation}
and $J_{\nu}$ and $Y_{\nu}$ correspond to Bessel functions of the first and second kinds, respectively, with an order of $\nu = 1/\left(2|1-\beta|\right)$.

\begin{figure}[htb!]
\begin{center}
\includegraphics[width=0.48\textwidth]{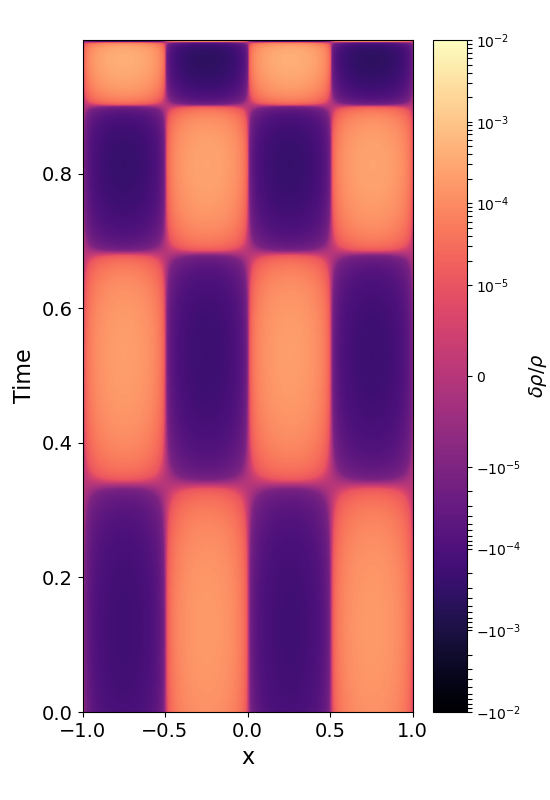}
\includegraphics[width=0.48\textwidth]{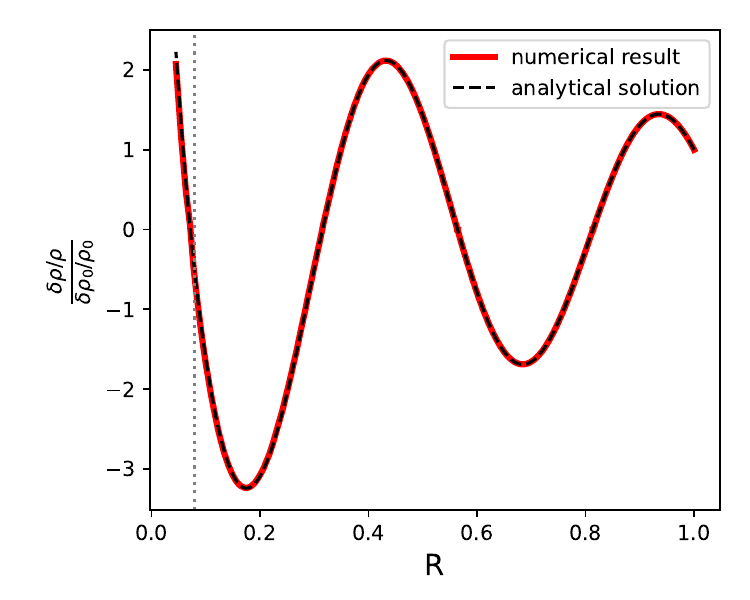}

\end{center}
\caption{Test case of purely horizontal sound wave with a nonuniform collapse profile. Upper panel: Space-time plot for vertically averaged relative density variation.  Lower panel: Relative amplitude of perturbation, $\delta \rho/\rho$, as a function of R from the simulation (solid red line), consistent with the analytical solution (dashed black line). The vertical dotted line marks the length scale, $L_{\mathrm{freezeout}}$, corresponding to the freeze-out regime. \label{fig:freezeout}}
\end{figure}

Additionally, given the relation $t_{\mathrm{bg}}=L/|\dot{L}|$ ($\beta \neq 1$), the collapsing system can evolve from the WKB regime ($\omega t_{\mathrm{bg}}\gg 1$) to the freeze-out regime ($\omega t_{\mathrm{bg}} \lesssim 1$), where the wave has a longer period than the background flow timescale, and can thus show a roughly static pattern. The corresponding length scale at freeze-out can be obtained by setting $\omega t_{\mathrm{bg}}=1$:

\begin{align} \label{eq:freezeout}
L_{\rm freezeout}=L_0\left( \frac{t_0}{t_c}\right)^{\beta} \left(\frac{|\beta|}{\omega_0t_0}\right)^{\beta/\left(1-\beta\right)}.
\end{align}

For a diagonally propagating sound wave, if the aspect ratio $b\neq 1$, the sound wave generates a shear flow perpendicular to the wave propagation direction in the local reference frame. This shear flow can be characterized by
\begin{align} \label{eq:shear}
&k_z\delta v_x-k_x \delta v_z = -\frac{2c_{\mathrm{s}}}{\mathcal{R}^{2}}k_x k_z \left(1-b^{-2} \right)\Re\left[X \exp \left(i \mathbf{k} \cdot \mathbf{x}\right)\right].
\end{align}
For a constant aspect ratio, $b$, this shear flow is not a physical phenomenon, but rather simply a consequence of the choice of a spatially anisotropic reference frame. 
However, with a varying aspect ratio during the collapse, the shear flow generated by the sound wave has a physical origin, as it cannot be removed by a time-independent rescaling of the anisotropic reference frame.

These aforementioned wave properties in the local model deserve careful testing numerically. Here, we tested the solutions for a horizontal sound wave with both a uniform and a general power law collapse profile separately, as well as the solution for diagonal sound waves.

\subsubsection{Horizontal sound waves with uniform collapse} \label{subsubsec:horizontal_wave}

We first considered a purely horizontally propagated traveling wave with a uniform collapse profile. This can also provide a comprehensive view of the wave behavior in the local box. We set up the following perturbation:
\begin{align}
&\delta v_{x0}  = A_0R_0^{-2} k_x \sin(k_x x), \\
&\delta v_{y0}  = \delta v_{z0} = 0,\\
&\delta \rho_0  =A_0\frac{\rho_0}{c_{s,H0}} R_0^{-2} k_x \sin(k_x x),
\end{align}
where $A_0$ is some (real) constant, and $c_{\mathrm{s},\mathrm{H0}}=c_\mathrm{s0}/R_0$ is the initial value of the effective horizontal sound speed defined in Sect. \ref{sec:equations}. This perturbation corresponds to wave amplitudes of $\left(X,\Pi\right)=\left( \frac{iA_0}{2c_\mathrm{s}}, \frac{A_0k_x}{2R_0}\right)$.

For the numerical test of the local model, we chose a collapse profile with $R_0 = 10.0L_0$, $U_{0} = 1.0c_{\mathrm{s}0}$, $L_{z0}=1.0L_0$, and $U_{R0}=0$, with a perturbation amplitude of $A_0 = 10^{-6}$, $k_x=16\pi$. The simulation box size was set as $L_{x,\mathrm{sim}}=0.5L_0$, $L_{z,\mathrm{sim}}=1.0L_0$, with a resolution of 256 cells per unit length.

The relative amplitude of gas density perturbation, $\delta \rho/ \rho$, was measured with respect to the background gas density, $\rho$, at each time point, which was obtained by averaging the gas density over the domain.  The space-time plot of the vertically averaged value of $\delta \rho/ \rho$ presented in the upper panel of Fig. \ref{fig:soundwave_h} shows the evolution of the wave pattern. The wave propagates rightward with a time-varying effective sound speed, $c_{\mathrm{s},\mathrm{H}}=c_{\mathrm{s}0}/\mathcal{R}$. As the collapse proceeds, the wave propagation is accelerated as $\mathcal{R}$ decreases, matching the overlaid curves of theoretical phase shift profiles over time. This time-varying effective sound speed is not physical in the global picture, but is instead due to our choice of a time-dependent reference frame in the local model.

Our choice of collapse profile results in the wave operating in the WKB regime. It is thus appropriate to compare the evolution of the wave amplitude to that expected from the WKB approximation solution. The lower panel of Fig. \ref{fig:soundwave_h} shows this comparison of the relative amplitude of density perturbation, $A_{\rho} = \max\left(\delta \rho / \rho \right)$. The numerical result of the $A_{\rho}$ evolution agrees well with the WKB approximation solution, $A_{\rho} = A_{\rho,0}\mathcal{R}^{-1/2}$.

Similarly, for a purely vertical sound wave, the corresponding length scale becomes $L_z$ instead of $\mathcal{R}$. We have also numerically tested the vertical sound wave and found that the effective vertical sound speed evolves following $c_{s,H}=c_{s0}/{L_z}$, and the relative amplitude of $\delta \rho$ follows $A_{\rho} = A_{\rho,0}L_z^{-1/2}$, which aligns with the theoretical expectations.

\subsubsection{Horizontal waves with nonuniform collapse} \label{subsubsec:non-uniform}
We next tested the horizontal sound waves with a more general power law collapse profile. In order to compare with the exact solution Eq. (\ref{eq:powexact1}-\ref{eq:powexact3}), we considered a collapse profile with $\beta=1/2$ and $t_\mathrm{c}=1$, i.e., $\mathcal{R} = {R}_0(1-t)^{1/2}$, and correspondingly initiated the perturbation of velocity and density as
\begin{equation} 
\begin{aligned} 
\delta v_{x0}  =A_0\frac{k_x}{{R_0}^2} s &Y_1(s_0) \cos(k_xx), \\ 
\end{aligned}
\end{equation}
\begin{equation} 
\begin{aligned} 
\delta v_{{y0}}  =\delta v_{{z0}}  =0, \\ 
\end{aligned}
\end{equation}
\begin{equation} 
\begin{aligned} 
\delta \rho_0 =A_0 \frac{\rho_0}{2 t_0 c_{\mathrm{s}0}^2}  &  Y_{0}(s_0)\sin(k_x x),
\end{aligned}
\end{equation}
where according to eq. (\ref{eq:sandt0}),  $s_0 = s(0) = \left(t_c/t_0\right)^{1-\beta}$, and $t_0$ can again be obtained from eq. (\ref{eq:sandt0}). $A_0$ is some real constant, and this perturbation corresponds to $A=0$, $B=A_0/2$ in eq. (\ref{eq:powexact1}-\ref{eq:powexact3}).
For the numerical test, we set $R_0=1.0L_0$, $L_{z0}=1.0L_0$, $U_{R0}=0$, with a perturbation of $A_0=10^{-5}$, $k_x=2\pi$, and a simulation box size of $L_{x,\mathrm{sim}}=2.0L_0$, $L_{z,\mathrm{sim}}=1.0L_0$, with a resolution of 256 cells per unit length.

Similar to Sect. \ref{subsubsec:horizontal_wave}, the relative amplitude of the gas density perturbation, $\delta \rho/ \rho$, was measured at each time point, where $\rho$ was obtained by averaging the gas density over the domain. The space-time plot of the vertically averaged relative gas density, $\delta \rho/ \rho$, displayed in the upper panel of Fig. \ref{fig:freezeout} shows that this perturbation leads to a horizontal standing wave. This wave exhibits phase reversal with each half wave period.  
We further measured the relative amplitude of the density perturbation by performing a Fourier transform to the wave pattern, and identified the amplitude of the dominant frequency component. The time evolution of the relative amplitude, $\delta \rho / \rho$,  is shown in the lower panel of Fig. \ref{fig:freezeout}, with the length scale corresponding to the freeze-out regime, $L_{\mathrm{ freezeout}}$ (see Eq. (\ref{eq:freezeout})), marked with a vertical dotted line. The result is in good agreement with the analytical solution in both the WKB regime ($\mathcal{R}>L_{\mathrm{freezeout}}$) and the freeze-out regime ($\mathcal{R}<L_{\mathrm{freezeout}}$). The L2 norm of the relative error for the relative amplitude, $\mathrm{err}(\delta \rho / \rho)$, defined in eq. (\ref{eq:err}) is on the order of $10^{-5}$ when the absolute value of the amplitude solution is much larger than $0$. When the solution of the amplitude is close to $0$ (i.e., close to the period of phase reversal), the relative error, $\mathrm{err}(\delta \rho / \rho)$, can be up to the order of $10^{-2}$, partly due to the division of a small analytical solution value, and partly due to inherent limitations and numerical inaccuracies that can arise near the phase reversal period. The slightly diffused wave pattern during the phase reversal can also be seen in Fig.~\ref{fig:freezeout}.

\begin{figure}
\begin{center}
\includegraphics[width=0.48\textwidth]{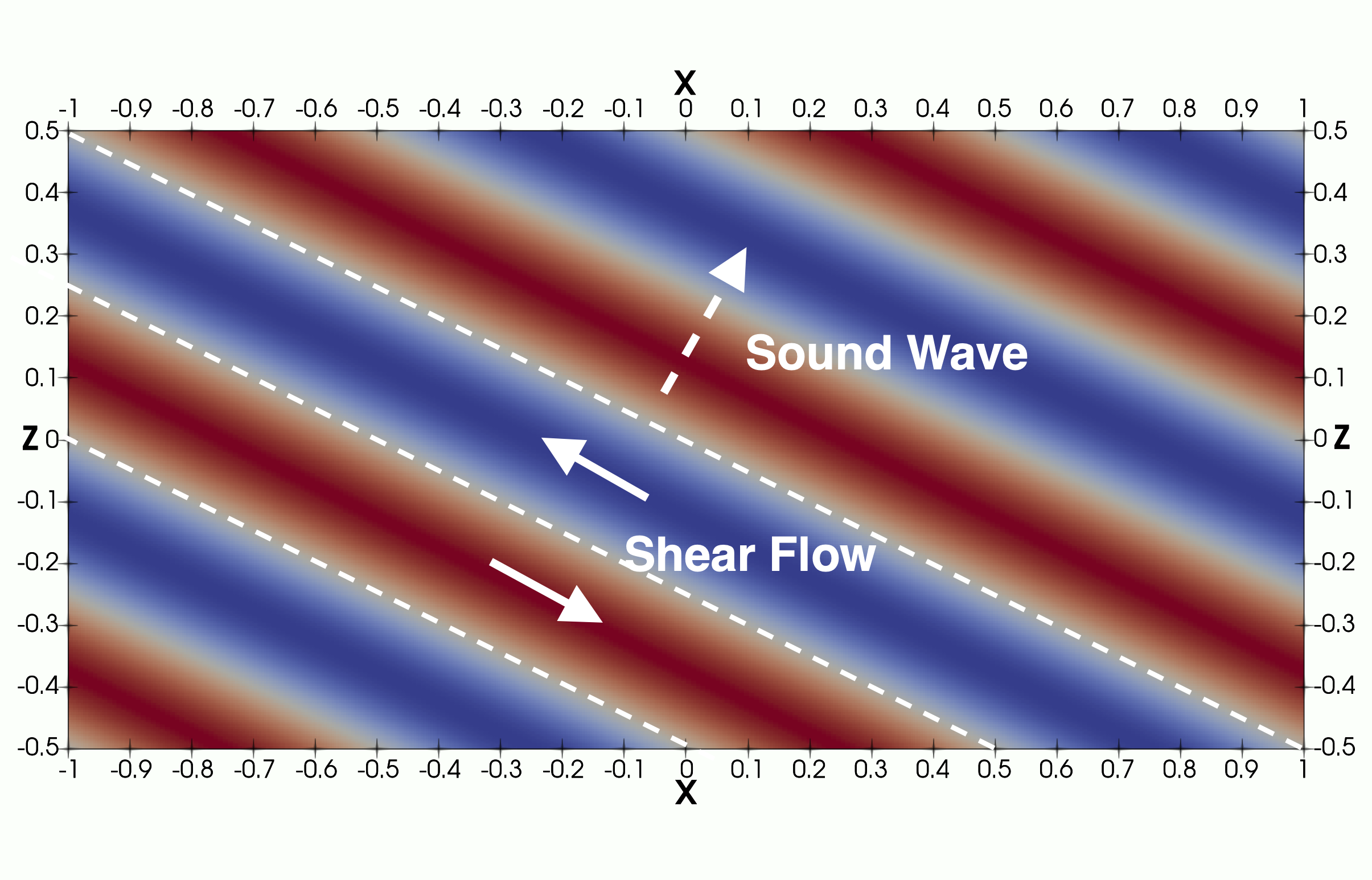}
\includegraphics[width=0.48\textwidth]{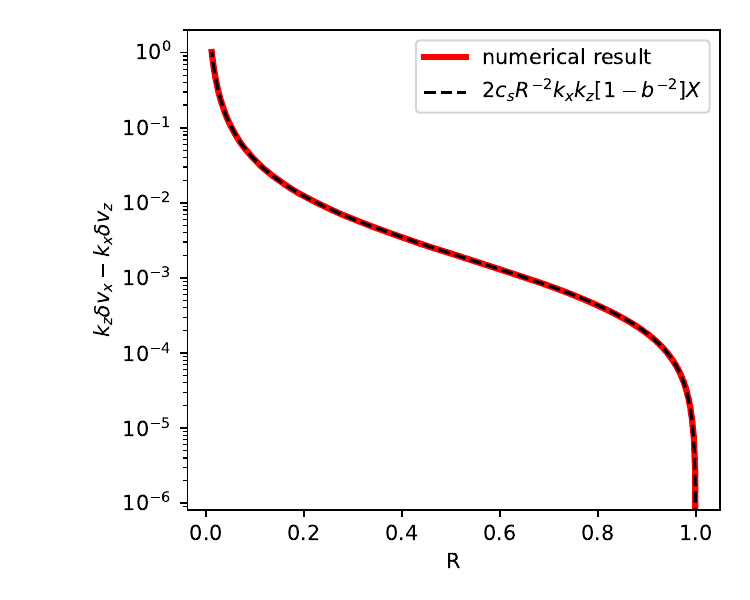}

\end{center}
\caption{
Test case for diagonal sound wave. 
Upper panel: Illustration of the diagonal sound wave and the shear flow generated. 
Lower panel: Amplitude of the shear flow generated by the wave as a function of R. The numerical result (solid red line) closely matches the analytical solution (dashed black line).
  \label{fig:diagonalwave}}
\end{figure}

\subsubsection{Diagonal sound waves} \label{subsec:diagwave}

Diagonally propagating sound waves may introduce numerical complexities, as they generate shear flows perpendicular to the direction of wave propagation. This is illustrated in the upper panel of Fig. \ref{fig:diagflow}, with the wave pattern characterized by the density.

We tested diagonal sound waves by generating density and velocity perturbations of
\begin{align}
&\delta v_{x0}  = A_0R_0^{-2} k_x \sin(k_x x + k_z z), \\
&\delta v_{y0}  = 0, \\
&\delta v_{z0} = A_0L_{z0}^{-2} k_z \sin(k_x x + k_z z),\\
&\delta \rho_0  =A_0\frac{\rho_0}{c_{\mathrm{s}0}} \sqrt{k_x^2R_0^{-2}+k_z^2L_{z0}^{-2}} \sin(k_x x + k_z z),
\end{align}
where $A_0$ is some (real) constant. This perturbation corresponds to an initial wave amplitude of $\left(X_0,\Pi_0\right)=\left( \frac{iA_0}{2c_{\mathrm{s}0}}, \frac{A_0}{2}\sqrt{k_x^2R_0^{-2}+k_z^2L_{z0}^{-2}}\right)$. Recalling Eq. (\ref{eq:hamilton1}), Eq. (\ref{eq:hamilton2}), and Eq. (\ref{eq:shear}), this pure diagonal sound wave generates a shear flow that can be characterized by
\begin{align}
\delta v_{\rm shear} &= k_z\delta v_x-k_x \delta v_z \\
&= -\frac{2c_\mathrm{s}}{\mathcal{R}^{2}}k_x k_z \left(1-b^{-2} \right)\operatorname{Re}\left[X \exp \left(i \mathbf{k} \cdot \mathbf{x}\right)\right],
\end{align}
and the dynamics of the system of $(X,\Pi)$ can be obtained from Eqs. (\ref{eq:hamilton1}) and (\ref{eq:hamilton2}).
Thus, the amplitude $X$ can be obtained by solving the equation
\begin{equation} \label{eq:solveivp}
\ddot{X}=-\omega^2(t) X,
\end{equation}
where the wave frequency, $\omega$, is given in Eq. (\ref{eq:omega}).

We numerically tested the diagonal sound wave by setting $R_0 = 1.0L_0$, $U_0=1.0c_{\mathrm{s}0}$, $L_{z0}=1.0L_0$, and $U_{R0}=0$, with a perturbation of $A_0=10^{-5}$, $k_x=2\pi$, $k_z=4\pi$, and a simulation box of $L_{x,\mathrm{sim}}=2.0L_0$, $L_{z,\mathrm{sim}}=1.0L_0$, with a resolution of 512 cells per unit length.

In the lower panel of Fig.~\ref{fig:diagonalwave}, we compare the amplitude of the shear flow generated by the diagonal sound wave in the numerical simulation with the analytical solution. The numerical result is obtained by calculating the shear flow, $\delta v_{\mathrm{shear}}$, in each grid, and performing a 2D Fourier transform of the shear flow map, followed by identifying the amplitude of the dominant frequency component. The analytical solution of the shear flow amplitude, $2{c_\mathrm{s}\mathcal{R}^{-2}}k_x k_z \left(1-b^{-2} \right)X $, was obtained by numerically solving for the wave amplitude, $X$, at each time step according to Eq. (\ref{eq:solveivp}), with the explicit Runge-Kutta integration method of order 5(4) \citep{Dormand80}. The numerical result and analytical solution of the shear flow amplitude are in good agreement. We further measured the L2 norm of the relative error for the shear strength, $\delta v_{\mathrm{shear}}$, in each grid, and got $\mathrm{err}(\delta v_{\mathrm{shear}})$ as defined in Eq.~(\ref{eq:err}) of the order of $10^{-6}$.

\begin{figure*}
\begin{center}
\includegraphics[width=0.48\textwidth]{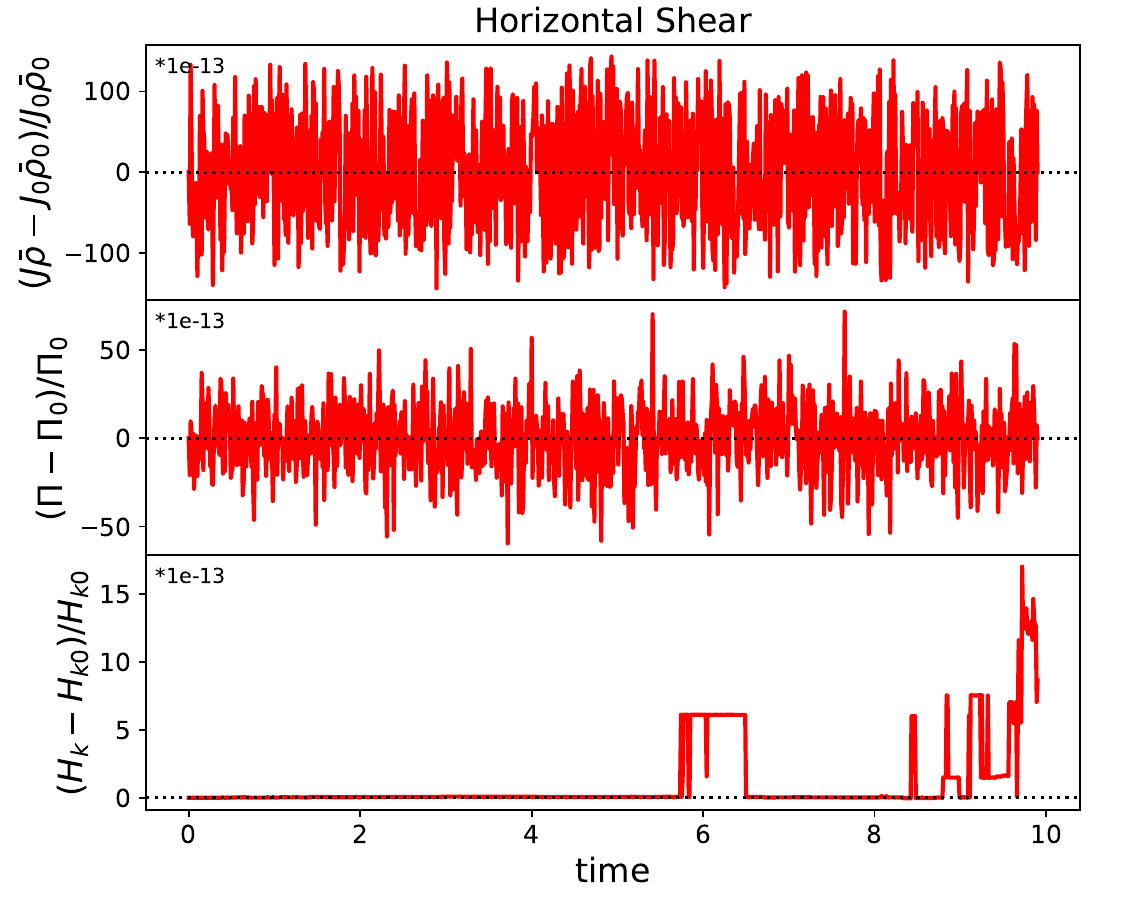}
\includegraphics[width=0.48\textwidth]{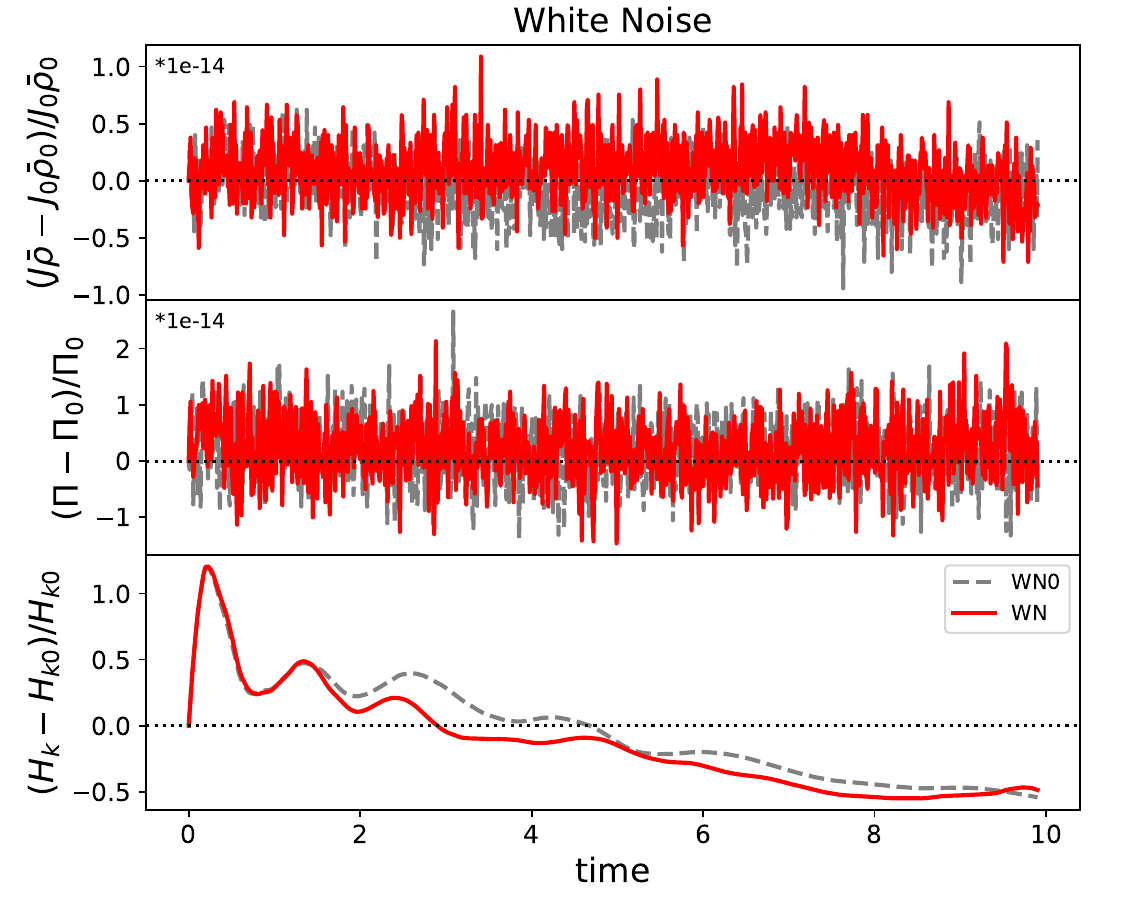}
\end{center}
\caption{ Conservation laws for tests of horizontal shear flow (left) and of white noise in the velocity field (right). For each case, the normalized variation in $J\bar{\rho}$ (left), $\Pi_i$ (middle), and $H_k$ (right) is presented as a function of time. 
For the horizontal shear flow test (see also Sect. \ref{subsec:3dshear}), the modified mass conservation law ($J\bar{\rho}$) and the conservation law for total momenta ($\Pi_i$) and for kinetic helicity $H_k$ are all satisfied to machine precision (note that for all of the panels, the y axis ticks are in units of $10^{-13}$).
For the case with white noise in the velocity field, the conservation laws for both mass, $J\bar{\rho}$, and total momenta, $\Pi_i$, are fulfilled to machine precision as well, but not for the kinetic helicity, $H_k$ (note that for the upper two panels, the y axis ticks are in units of $10^{-14}$).
For comparison, a simulation with the same setup of white noise, but without collapse (i.e., $U_0=U_{R0}=0$) was performed (``WN0", dashed grey lines, see also the legend), which gives consistent results to WN. We thus anticipate that the deviation from conservation law for $H_k$ is likely due to intrinsic characteristics of \textsc{Athena++}, which in certain situations is not strictly vorticity-conserving.   \label{fig:conservation}}
\end{figure*}   

\subsection{Conservation laws} \label{subsec:conservation}

The continuity equation of the local model, Eq. (\ref{eq:mass}), gives the conservation law for the total mass,
\begin{equation}
\begin{aligned} M & =\iiint \rho \mathrm{d}V = \iiint \rho J \mathrm{d}x \mathrm{d}y \mathrm{d}z,\end{aligned}
\end{equation}
where $J \equiv \mathcal{R}^2L_z$ is the Jacobian determinant in the local coordinate system. 

The momentum equation of the local model, Eq. (\ref{eq:momx} - \ref{eq:momz}), gives the conservation law for the total momenta,
\begin{equation}
\begin{aligned} \Pi_i & =\iiint \rho \mathcal{V}_i \, \mathrm{d}V = \iiint \rho \mathcal{V}_i J \,  \mathrm{d}x \mathrm{d}y \mathrm{d}z,\end{aligned}
\end{equation}
where $\mathcal{V}_i$ is the covariant velocity component in each direction,
\begin{equation}
\mathcal{V}_x =  v_x \mathcal{R}^2, \mathcal{V}_y =  v_y \mathcal{R}^2, \mathcal{V}_z =  v_z L_z^2.
\end{equation}

Another conserved quantity in the local model is the kinetic helicity, $H_k$, of the relative flow, which is a quantity associated with the vorticity. It is defined as
\begin{equation}
\begin{aligned} H_k & =\iiint \mathcal{V}_i  {\omega^i}  \,  \mathrm{d} V  =\iiint \mathcal{V}_i   {\omega^i} J \,  \mathrm{d} x \mathrm{d}y \mathrm{d} {z},\end{aligned}
\end{equation}
where $\omega^i$ is the fluid (contravariant) vorticity, 
\begin{equation}
\begin{aligned}   {\omega^i}    = \varepsilon^{i j k} \partial_j \mathcal{V}_k ,\end{aligned}
\end{equation}
and $\varepsilon^{i j k}$ is the volume element of the local model. Details and derivations of the conservation laws in the local model are discussed in \S 2.5 of \citetalias{Lynch23}. 

In order to validate these conservation laws in the local box, we performed a simulation run with a uniform background density, $\rho_0=1$, introducing white noise perturbations in the velocity field, $v_x, v_y$, and $v_z$, with a relative amplitude of  $10^{-6}$. We chose a collapse profile with $R_0=10.0$, $U_0=-1.0c_{\mathrm{s}0}$, $L_{z0}=10.0$, and $U_{R0}=-0.1$, and set the simulation box size as $L_{x,\mathrm{sim}}=L_{y,\mathrm{sim}}=0.5$, $L_{z,\mathrm{sim}}=1.0$, with a resolution of 32 cells per unit length. This test run is labeled as ``WN." For comparison, we also performed a simulation with the same setup for the white noise, but without collapse (i.e., $U_0=U_{R0}=0$), which is labeled as ``WN0." 
In addition, the result of the horizontal shear flow test (see Sect. \ref{subsec:3dshear}) was also used to examine the conservation laws.

We measured the volume-averaged density, $\bar{\rho}$ , total momenta, $\Pi$, and kinetic helicity, $H_{\rm k}$, in the simulations, and present in Fig. \ref{fig:conservation} the normalized variation in $J\bar{\rho}$ , $\Pi$, and $H_{\rm k}$ as a function of time. 
Our results demonstrate that the modified mass conservation law ($J\bar{\rho}$) and the conservation law for total momenta ($\Pi_i$) are both satisfied to machine precision. 
The kinetic helicity conservation, although preserved to machine precision for the shear flow test, exhibits significant deviation for the white noise simulations. 
This discrepancy persists in both WN and WN0 cases, suggesting that the lack of conservation is not a consequence of the integration of the collapsing box. Instead, it is likely due to intrinsic characteristics of the \textsc{Athena++} framework, which in certain situations is not strictly vorticity conserving, likely because the higher-order Godunov scheme generally introduces a degree of numerical diffusion \citep[see e.g.][]{Seligman17}. In the white noise case in which vorticity is dominated by fine-scale structures, there is increased numerical dissipation at small scales because the slope limiter is expected to lead to the nonconservation of $H_{\rm k}$. 
For an adiabatic EoS, the entropy conservation law is also validated; this is further presented in App. \ref{sec:Eint}.

\subsection{Convergence study} \label{subsec:convergence}

\begin{figure}
\begin{center}
\includegraphics[width=0.48\textwidth]{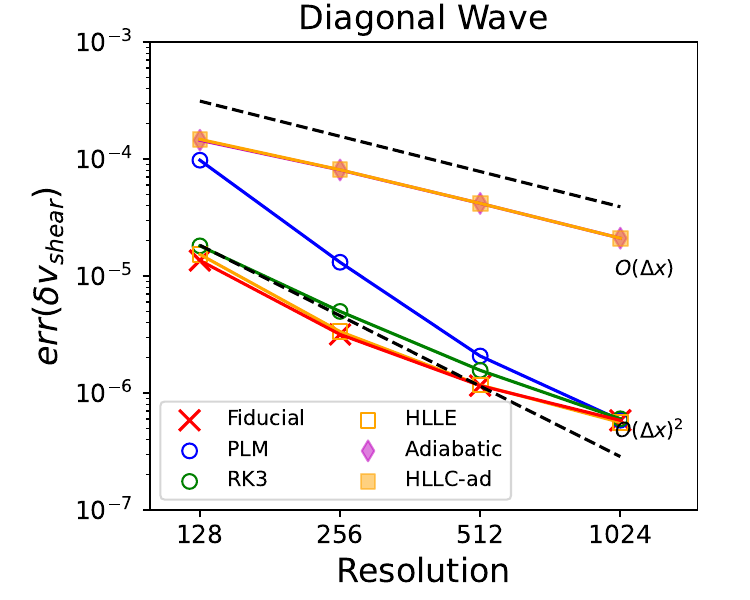}

\end{center}
\caption{Time-averaged L2 norm relative error for diagonal sound wave test, as a function of simulation resolution (cells per unit length). Overall, convergence between the first and second orders is achieved (see the dashed lines for reference).} \label{fig:convergence}
\end{figure}

We performed a convergence study with the diagonal wave (Sect. \ref{subsec:diagwave}) test, as it shows the maximum possible error. Basic simulation setups such as wave properties and box sizes are the same as presented before. Fiducially, simulations were performed with Roe's solver, the VL2 integrator, PPM reconstruction, and an isothermal EoS. The CFL number in the convergence study was fixed to $0.2$. However, we validated the numerical stability of our code for CFL numbers up to 0.4 for VL2 and 0.8 for RK3. 
We further explored the effects of different Riemann solvers, time integrators, spatial reconstruction methods, and the EoS on the numerical performance of the code. We also varied the spatial resolution of our simulation in units of cells per unit length.
The simulation setups for test runs that we performed for the convergence study are presented in Table \ref{tab:sims}.
Figure \ref{fig:convergence} shows the time-averaged L2 relative error, $\mathrm{err}(\delta v_{\mathrm{shear}})$, as a function of resolution.
\begin{table} 
\caption{Simulation setup for the convergence study. \label{tab:sims}}
\centering
\begin{tabular}{| c|c|c|c|c|}
 \hline
 Sim Label & Solver & Integrator & Reconst.&EoS\\
 \hline \hline
 Fiducial   & Roe    &VL2&   PPM &Isothermal\\ 
 HLLE    &\bf{HLLE} & VL2&  PPM &Isothermal\\
 RK3 & Roe & \bf{RK3}&  PPM &Isothermal\\
 PLM &   Roe  & VL2&\bf{PLM} &Isothermal\\  
 Adiabatic&   Roe& VL2&PPM &\bf{Adiabatic}\\ 
 HLLC-ad & \bf{HLLC}& VL2&PPM &\bf{Adiabatic}\\
 \hline 
\end{tabular}
\tablefoot{Variations compared to the fiducial simulation highlighted in bold. ``Reconst.'' stands for reconstruction.}
\end{table}

Our results demonstrate that our code generally achieves convergence between the first and second orders for the diagonal wave test. 
The performances for test cases HLLE and RK3 remain generally similar to the fiducial case, while a lower order of spatial reconstruction (PLM) leads to a lower accuracy at lower resolutions.
These tests with an isothermal EoS achieve second-order convergence at lower resolutions, with a tentative trend toward the first order at high resolutions. This trend is likely attributed to time-dominated errors at higher resolutions, as our source term incorporation is first-order in time. We also explored the higher-order method to integrate source terms with the VL2 integrator, which was not as efficient as our first-order method, showing a comparable or slightly higher error, and reduced stability for adiabatic flows. Therefore, we chose to use our first-order method. The source term can also be incorporated into higher-order methods with Runge-Kutta integrators. We leave the development of higher-order methods to future work. 

The simulation runs with the adiabatic EoS (for both the Roe and the HLLC solvers) exhibit systematically higher ($\sim$ an order of magnitude) error levels, with first-order convergence. This pattern is likely attributed to the complexities that the adiabatic EoS introduces to the system, including the additional energy equation to be solved. Additionally, we note that our simulations may encounter stability issues in scenarios involving supersonic adiabatic flows with added perturbations (white noise). Future work will address these challenges and improve the handling of adiabatic processes within our simulations.

\section{Discussion}
\label{sec:discussion}

The complete series of tests presented in Sect. \ref{sec:tests}, including the challenging diagonal wave test that induces shear flows, demonstrates the robustness and stability of our \textsc{Athena++} code implementation, confirming its capability to handle complex local model simulations effectively.
\textsc{Athena++} has also been used to incorporate similar local models; namely, the uniform expanding box for studying stellar winds \citep{Squire20} and the anisotropic expanding box implemented in the MHD-PIC module \citep{Sun23}. 
These previous numerical implementations have concentrated on  investigating MHD characteristics and the behavior of magnetized waves. Our study focuses on a comprehensive examination of the hydrodynamic aspects of the expanding and collapsing box.
Moreover, while our code is adaptable to scenarios akin to the aforementioned expanding cases, here we focus on collapsing systems. 
This is motivated not only by our interest in probing protostellar collapse, but also by the fact that collapsing environments - with their intrinsic amplification of density, velocity, vorticity, and energy - are potentially more abundant in instabilities. Indeed, our observations of the growth of shear flows and wave-induced shear underline the possibility of hydrodynamic instabilities during protostellar collapse. Moreover, related previous work \citep{Robertson12} also presented an increase in turbulent velocities (``adiabatic heating'') of contracting gas. It is thus both numerically and scientifically important to conduct a thorough investigation and validation of this local collapsing box model before delving into MHD studies.

The magnetic field plays an important role in protostellar collapse, as the core must lose angular momentum during the collapse \citep[see e.g. review of ][]{McKee07}, and magnetic braking has been considered as one of the main mechanisms driving angular momentum transfer during core collapse and forming protoplanetary disks with realistic properties \citep[e.g.][etc.]{Allen03b,Price07,Hennebelle08a,Commercon11,Masson16,Machida16,Wurster16,Hennebelle20}. 
The extension to MHD for the theoretical model of the expanding or collapsing box will likely follow a similar approach to the MHD eccentric shearing box \citep{Ogilvie14}. The numerical implementation will presumably be similar to the approach described in this paper. However, magnetic fields in collapse processes may break spherical symmetry \citep{Galli93,Hennebelle01,Galli06}, and thus may require further modifications of the local model. One should also be careful with the treatment of the background magnetic field gradient, in order to avoid spurious instabilities \citep{Latter17}.

The local deviation from the spherical symmetry, such as that caused by the magnetic field discussed above, leads to the need for further generalization of this local box toward more realistic collapse profiles. A weak rotation in the global picture can be incorporated by simply using shear-periodic boundary conditions, whereas a strong rotation requires modification of the local model.
Another factor to be considered is physical collapse profiles. As is shown in Sect. \ref{subsubsec:non-uniform}, a time-dependent collapse profile can transition between different regimes, including from subsonic to supersonic flows, or from WKB to freeze-out wave regimes. It is thus necessary to study the effect of various collapse profiles, such as the typical profile of  \citet{Shu77}.

Dust grains are the building blocks of planets.
Recent surveys of protoplanetary disks across different stages of disk evolution \citep[e.g.][]{Pascucci16,Barenfeld16,Ansdell18,Long18,Cieza19,Tychoniec20,Tobin20,VanderMarel21} show that the available solid mass to form planets is rapidly depleted during the early stage of disk evolution, likely indicating an early start of planet formation. 
Observational\citep{Pagani10,Kataoka15,Galametz19,Valdivia19} and theoretical \citep[e.g.][etc.]{Ormel09,Hirashita09,Guillet20,Bate22}{} works both suggest that dust grains are growing in collapsing protostellar cores, significantly prior to the protoplanetary disk phase. 
With further incorporation of dust, our local collapse simulation will be capable of resolving and investigating dust dynamics and their interactions with gas, especially with potential instabilities and other local phenomena, which have a rich potential to concentrate dust during the collapse. 
Moreover, the interaction between dust and gas drag force might give rise to additional dust-related instabilities.
One example, the streaming instability \citep{Youdin05}, has already been proven to be a promising avenue for planetesimal formation in protoplanetary disks, and has recently been extended to 
a family of instability caused by dust-gas interaction \citep[resonance drag instability,][]{Squire18}, which could potentially operate to concentrate dust in the context of protostellar collapse.  
These dust concentration mechanisms in the initial collapse phase might significantly expedite the onset of planet formation, potentially starting before the complete disk formation. This aligns with the aforementioned observational indicators of early planet formation in very young stellar systems.
The implementation of dust for the current model is likely similar to the method presented in this paper. However, the relative drift between the background flows of gas and dust has to be implemented in an ad hoc way, similar to the approach of the shearing box \citep[e.g.,][]{Bai10}.

\section{Conclusion} 
\label{sec:conclusion}

We implemented the local collapsing box model of \citetalias{Lynch23} into \textsc{Athena++}, and performed benchmark tests for the behavior of nonlinear and linear solutions in the local model. 
This model features a spatially uniform time-dependent Cartesian geometry, allowing (shear-)periodic boundary conditions. The key points of this study are:

\begin{itemize}
    \item We numerically implemented the local model by modifying the Riemann solver with a rescaled anisotropic effective sound speed and an adapted energy equation, together with operator split-source terms.
    \item Our benchmark test of horizontal shear flow shows that the shear is amplified as the collapse proceeds, in agreement with the analytical solution to machine precision. This corresponds to a zonal flow in the global picture, and the flow amplification reflects angular momentum conservation. Similar phenomena are verified with tests of the elevator flow and the diagonal flow, which also have corresponding physical interpretations.
    \item Our benchmark tests of sound waves match the theoretical predictions of wave speed and amplitude evolution over time in the local model, for both uniform and general power-law collapse profiles. Additionally, a diagonal wave generates a shear flow, which could pose numerical challenges that are well handled by our implementation. The growth and evolution of this shear flow are also rigorously tested in our simulation, with a typical relative error level of $10^{-4} - 10^{-6}$. 
    \item The density, momentum, and vorticity in the local model increase as the collapse proceeds. The corresponding (modified) conservation laws for mass and total momenta are validated in our code.
    \item Our code generally achieves convergence between the first and second orders for the diagonal wave-generated shear flow test.  Further improvement in performance with an adiabatic EoS is necessary for future work.
\end{itemize}

This local collapsing box implementation in \textsc{Athena++} is not only crucial for probing the local phenomena and hydrodynamic instabilities of protostellar collapse, but also holds potential for future adaptations to include MHD processes and dust dynamics with more realistic collapse profiles. Its ability and potential to simulate local behaviors and instabilities and the interplay of various physical factors positions it as a valuable tool for advancing studies of protostellar collapse and disk formation, as well as the earliest stages of planet formation. The code will be available upon request for collaboration and will be publicly released shortly after additional benchmark tests and improvements.

\begin{acknowledgements}
We thank Francesco Lovascio, Benoît Commerçon, Xuening Bai, Xiaochen Sun, and Lile Wang for constructive discussions. We thank the anonymous referee for a thorough and helpful report. ZX, EL and GL acknowledge the support of the European Research Council (ERC) CoG project PODCAST No. 864965.
\end{acknowledgements}




\bibliographystyle{mnras}
\bibliography{collapsing_box} 

\onecolumn

\begin{appendix}

\section{Adiabatic equation of state}
\label{sec:adiabatic}

The eigenvalues and eigenmatrices in the conserved variables with an adiabatic EoS are needed to construct the fluxes of the conserved variables in the Roe's solver. These quantities are provided hereafter

\begin{equation}
\lambda=\left(v_1-C, v_1, v_1, v_1, v_1+C\right) \text {, }
\end{equation}
\begin{equation}
\mathscr{L}=\left[\begin{array}{ccccc}
N_a\left(\gamma^{\prime} v^2 / 2+v_1 C\right) & -N_a\left(\gamma^{\prime} v_1+C\right) & -N_a \gamma^{\prime} v_2 & -N_a \gamma^{\prime} v_3 & N_a \gamma^{\prime} \\
-v_2 & 0 & 1 & 0 & 0 \\
-v_3 & 0 & 0 & 1 & 0 \\
1-N_a \gamma^{\prime} v^2 & \gamma^{\prime} v_1 / C^2 & \gamma^{\prime} v_2 / C^2 & \gamma^{\prime} v_3 / C^2 & -\gamma^{\prime} / C^2 \\
N_a\left(\gamma^{\prime} v^2 / 2-v_1 C\right) & -N_a\left(\gamma^{\prime} v_1-C\right) & -N_a \gamma^{\prime} v_2 & -N_a \gamma^{\prime} v_3 & N_a \gamma^{\prime}
\end{array}\right],
\end{equation}
\begin{equation}
\mathscr{R}=\left[\begin{array}{ccccc}
1 & 0 & 0 & 1 & 1 \\
v_1-C & 0 & 0 & v_1 & v_1+C \\
v_2 & 1 & 0 & v_2 & v_2 \\
v_3 & 0 & 1 & v_3 & v_3 \\
H-v_1 C L_1^2 & v_2 L_2^2 & v_3 L_3^2 & v^2 / 2 & H+v_1 C L_1^2
\end{array}\right],
\end{equation}
where $N_a \equiv 1 /\left(2 C^2\right)$ and $\gamma^{\prime} \equiv \gamma-1$. In addition, in the local model we have redefined 
\begin{equation}
C = c_s/L_1 \text{,}
\end{equation}
\begin{equation}
v^2= v_1^2 L_1^2 + v_2^2 L_2^2 + v_3^2 L_3^2 \text{,}
\end{equation}
and
\begin{equation}
H = \mathcal{E}_{\text {rell }} + p/\rho. 
\end{equation}
The subscripts 1,2,3 in $L_1$,$L_2$,$L_3$ and $v_1$, $v_2$, $v_3$ denote the corresponding quantities along the three axes of a Cartesian coordinate system, with an order that respects the right-hand orientation, i.e., in the order of $(x,y,z)$, $(y,z,x)$, or $(z,x,y)$. $L_1$, $L_2$, $L_3$ are the characteristic length scales on the corresponding directions, which is $\mathcal{R}$ on $x$ and $y$ directions, and $L_z$ on the $z$ direction.

The eigenvalues and eigenmatrices in the primitive variables are needed for the reconstruction algorithm. These quantities are 
\begin{equation}
\lambda=\left(v_1-C, v_1, v_1, v_1, v_1+C\right) \text {, }
\end{equation}
\begin{equation}
\mathscr{L}=\left[\begin{array}{ccccc}
0 & -\rho/(2C) & 0 & 0 & 1/(2C^2L_1^2) \\
1 & 0 & 0 & 0 & -1/(C^2L_1^2) \\
0 & 0 & 1 & 0 & 0 \\
0 & 0 & 0 & 1 & 0 \\
0 & \rho/(2C) & 0 & 0 & 1/(2C^2L_1^2)
\end{array}\right],
\end{equation}
\begin{equation}
\mathscr{R}=\left[\begin{array}{ccccc}
1 & 1 & 0 & 0 & 1 \\
-C/\rho & 0 & 0 & 0 & C/\rho \\
0 & 0 & 1 & 0 & 0 \\
0 & 0 & 0 & 1 & 0 \\
C^2L_1^2 & 0 & 0 & 0 & C^2 L_1^2
\end{array}\right].
\end{equation}

\section{Derivation of the CFL condition}
\label{sec:cfl}
In addition to the standard CFL condition, we further consider the effect of the time-varying sound speed on the stability condition.
Consider a harmonic oscillator equation 
\begin{equation}
\ddot{y} + \omega^2\left(t\right) y = 0, 
\end{equation}
with a time-varying frequency $\omega\left(t\right)=\omega_0\left(1+\delta \Delta t\right)$, where $\omega_0$ is the initial frequency, and $\delta = \dot{\omega_0} / \omega_0$. Applying leapfrog FDE
\begin{equation}
\frac{y^{n+1}-2y^n+y^{n-1}}{\Delta t^2} + \omega_0^2\left(1+\delta \Delta t\right)^2 y^n = 0,
\end{equation}
and look for solutions of the form $y^n = y_0 e^{i \omega t}$, leading to
\begin{equation}
e^{i\omega \Delta t} - 2 + e^{-i\omega \Delta t} + \omega_0^2\Delta t^2\left(1+\delta \Delta t\right)^2 = 0,
\end{equation}
which simplifies to
\begin{equation}
\sin^2\left(\frac{\omega \Delta t}{2}\right) = \left(\frac{\omega_0\Delta t}{2}\right)^2 \left(1+\delta \Delta t\right)^2.
\end{equation}
This yields solutions of 
\begin{equation}
\omega = \frac{2}{\Delta t} \arcsin\left[\pm \frac{\omega_0\Delta t}{2} \left(1+\delta \Delta t\right)\right],
\end{equation}
and the stability condition is thus
\begin{equation}
\left| \frac{\omega_0\Delta t}{2} \left(1+\delta \Delta t\right) \right| \leq 1.
\end{equation}
Let $x=\omega_0\Delta t>0$, $\beta = \frac{\delta}{\omega_0}$, the stability condition then becomes
\begin{equation} \label{eq:stability}
\left|\beta x^2 + x\right| -2 <0.
\end{equation}

For $\beta=0$, the stability condition becomes $x<2$, i.e., $\Delta t < 2\omega_0^{-1}$, which goes back to the original CFL condition. It is convenient to write the CFL condition in the form of $\Delta t < 2\omega_0^{-1}\mathrm{v}(\beta)$, where $\mathrm{v}(\beta)$ the additional ``correction'' term that reflects the time-varying sound speed, and here we have $\mathrm{v}(\beta)=1$.

For $\beta >-1/8$, the stability condition eq. (\ref{eq:stability}) yields 
\begin{equation}
x < \frac{-1+\sqrt{1+8\beta}}{2\beta},
\end{equation}
thus 
\begin{equation}
\mathrm{v}(\beta) = \frac{-1+\sqrt{1+8\beta}}{4\beta}.
\end{equation}

For $\beta <-1/8$, the stability condition eq. (\ref{eq:stability}) yields 
\begin{equation}
x < \frac{-1-\sqrt{1-8\beta}}{2\beta},
\end{equation}
thus 
\begin{equation}
\mathrm{v}(\beta) = \frac{-1-\sqrt{1-8\beta}}{4\beta}.
\end{equation}

\section{Conservation of entropy} \label{sec:Eint}
For an adiabatic EoS in the absence of shocks, the local model leads to the conservation law for the following quantity

\begin{equation}
\iiint \varepsilon \rho^{2-\gamma} J dx dy dz.
\end{equation}

For a system with uniform density, this leads to conservation for the following quantity related to the entropy
\begin{equation}
K \equiv U_{int} M^{-\gamma} J^{\gamma},
\end{equation}
where $M$ is the total mass in the system, and
\begin{equation}
U_{int} \equiv \iiint \rho \varepsilon dx dy dz,
\end{equation}
the total internal energy in the system is thus $U_{int}/J$.

In order to validate the conservation law of entropy in the local collapsing box, we present the results for the horizontal shear flow test runs (see Sect. \ref{subsec:3dshear}) with adiabatic EoS, using Roe and HLLC solvers, for examining the conservation law. These runs are marked with the same labels as in Table \ref{tab:sims}. 

For both the Roe and HLLC solvers, $K$ is conserved to the order of $10^{-10}$. 

\begin{figure*}
\begin{center}
\includegraphics[width=0.48\textwidth]{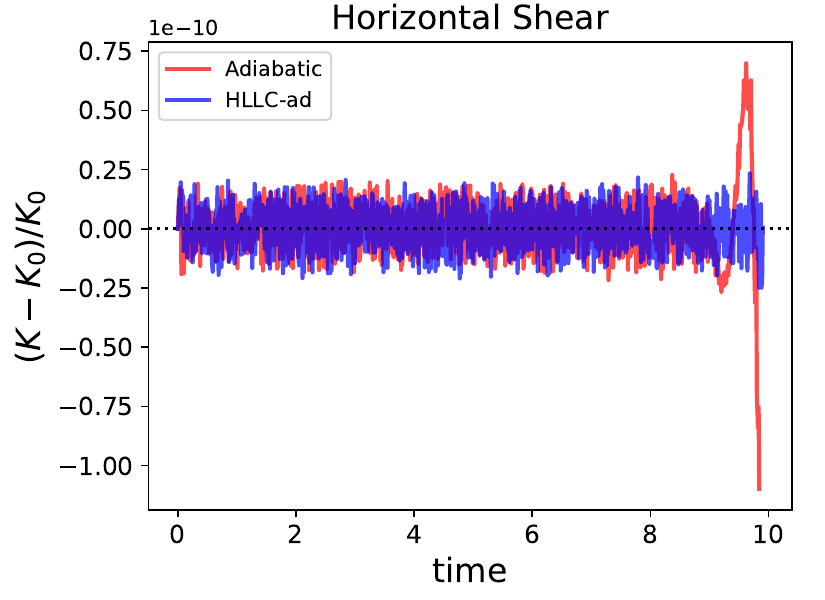}
\end{center}
\caption{ Conservation law of entropy with adiabatic EoS for tests of horizontal shear flow. For each case, the normalized variation of $K$ are presented as a function of time. The results with the Roe's solver are presented with red curves, while the results with the HLLC solver are presented in blue.
The conservation law for entropy is satisfied to the order of $10^{-10}$.   \label{fig:conservation_ad}}
\end{figure*}

\end{appendix}


\label{lastpage}
\end{document}